\documentclass[11pt,epsf,reqno]{amsart} 

\usepackage{times}

\usepackage{amssymb}
\usepackage{amsxtra}
\usepackage{amsmath}

\usepackage{verbatim}
\usepackage{epsfig}
\usepackage{setspace}

\newtheorem{theorem}{Theorem}[section]
\newtheorem{definition}{Definition}[section]
\newtheorem{lemma}[theorem]{Lemma}
\newtheorem{proposition}[theorem]{Proposition}
\newtheorem{corollary}[theorem]{Corollary}
\newtheorem{example}{Example}[section]
\newtheorem{conjecture}{Conjecture}[section]

\def\d{\delta}
\def\bd{\bar{\delta}}
\def\gm{\gamma}

\def\k{\kappa}
\def\l{\lambda}

\def\G{\Gamma}

\def\R{{\mathbb R}}

\def\F{{\mathbb F}}

\def\b{{\mathbf b}}
\def\c{{\mathbf c}}
\def\g{{\mathbf g}}
\def\h{{\mathbf h}}

\def\s{{\mathbf s}}

\def\bv{{\mathbf v}}

\def\x{{\mathbf x}}
\def\y{{\mathbf y}}

\def\0{{\mathbf 0}}
\def\1{{\mathbf 1}}
\def\O{{\mathbf O}}
\def\cC{{\mathcal C}}
\def\cD{{\mathcal D}}
\def\cE{{\mathcal E}}
\def\cF{{\mathcal F}}
\def\cG{{\mathcal G}}
\def\cH{{\mathcal H}}
\def\cI{{\mathcal I}}
\def\cJ{{\mathcal J}}

\def\cM{{\mathcal M}}
\def\cS{{\mathcal S}}
\def\cT{{\mathcal T}}

\def\cV{{\mathcal V}}

\def\cX{{\mathcal X}}

\def\tb{\widetilde{\b}}

\def\hc{\hat{c}}
\def\hbc{\hat{\c}}

\def\mfC{{\mathfrak C}}
\def\mfD{{\mathfrak D}}
\def\mfG{{\mathfrak G}}
\def\mfR{{\mathfrak R}}

\def\sfv{{\sf v }}

\def\supp{\mbox{\textsf{supp}}}
\def\rank{\mbox{\textsf{rank}}}
\def\rref{\mbox{\textsf{rref}}}

\def\2s{{\oplus}_2}
\def\3s{{\oplus}_3}
\def\d3s{\,\overline{\oplus}_3\,}
\def\bar{\overline}

\def\shorten{\setminus\!}
\def\delete{\shorten}

\def\pl{\parallel}
\def\la{\langle}
\def\ra{\rangle}

\title{A Decomposition Theory for Binary Linear Codes}
\thanks{This work was supported by a Discovery Grant
from the Natural Sciences and Engineering Research Council (NSERC), 
Canada.}
\author{Navin Kashyap} 

\thanks{N.\ Kashyap is with the Department of Mathematics and Statistics,
Queen's University, Kingston, ON K7L 3N6, Canada.
Email: \texttt{nkashyap@mast.queensu.ca}}

\renewcommand{\markboth}[2]
{\renewcommand{\leftmark}{\scshape{#1}}\renewcommand{\rightmark}{\scshape{#2}}}

\begin{document}

\renewcommand{\thefootnote}{\arabic{footnote}}
\setcounter{footnote}{0}
\begin{abstract}
The decomposition theory of matroids initiated by Paul Seymour in 
the 1980's has had an enormous impact on research in matroid theory.
This theory, when applied to matrices over the binary field,
yields a powerful decomposition theory for binary linear codes. In this
paper, we give an overview of this code decomposition theory, and
discuss some of its implications in the context of the recently
discovered formulation of maximum-likelihood (ML) decoding of a binary
linear code over a discrete memoryless channel as a linear programming
problem. We translate matroid-theoretic results of Gr\"otschel and
Truemper from the combinatorial optimization literature to give
examples of non-trivial families of codes for which the ML decoding
problem can be solved in time polynomial in the length of the code.
One such family is that consisting of codes $\cC$ 
for which the codeword polytope is identical to the 
Koetter-Vontobel fundamental polytope 
derived from the entire dual code $\cC^\perp$.
However, we also show that such families of codes are not 
good in a coding-theoretic sense --- either their dimension or 
their minimum distance must grow sub-linearly with codelength.
As a consequence, we have that decoding by linear programming,
when applied to good codes, cannot avoid failing occasionally
due to the presence of pseudocodewords.
\end{abstract}

\date{\today}
\maketitle



\section{Introduction}

Historically, the theory of error-correcting codes has benefited greatly 
from the use of techniques from algebra and algebraic geometry. Combinatorial
and graph-theoretic methods have also proven to be useful in the 
construction and analysis of codes, especially within the last ten
years since the re-discovery of low-density parity-check codes. However,
one area of mathematics that has, surprisingly, only played a 
relatively minor role in the development of coding theory is the 
field of matroid theory. The reason this is surprising is that, 
as anyone with even a basic understanding of the two fields would
realize, coding theory and matroid theory are very closely related. 
The former deals with matrices over a finite field $\F$, and these objects
are also of fundamental importance in the latter, where they go 
under the name of $\F$-representable matroids.

The connection with matroid theory has not gone unnoticed among 
coding theorists. Indeed, as far back as 1976, Greene \cite{greene}
noticed that the Tutte polynomial of a matroid, 
when specialized to a linear code $\cC$,
was equivalent (in a certain sense) to the homogeneous weight-enumerator 
polynomial $W_{\cC}(x,y) = \sum_i A_i x^i y^{n-i}$, where $A_i$
is the number of words of weight $i$ in $\cC$, and $n$ is
the length of $\cC$. The MacWilliams identities are then a special
case of a known identity for the Tutte polynomial 
\cite[Chapter~4]{cameron}. The connection with matroids was also exploited
by Barg \cite{barg} to derive MacWilliams-type identities for 
generalized-Hamming-weight enumerators of a code.

However, aside from such use of tools from matroid theory to re-derive 
results in coding theory that had already been proved by other means, 
each field seems to have had little impact on the other. Matroid theory
has derived its inspiration largely from graph theory, and its most
successful applications have traditionally been in the areas of 
combinatorial optimization and network flows. Very recently, matroid
theory has found new applications in the rapidly evolving field
of network coding \cite{dougherty}. 

On the other hand, coding theory (by which we mean the theory of 
error-correcting codes, in contrast to the theory of network coding) 
has been largely unconcerned with developments in 
combinatorial optimization, as the fundamental problems
in the former seemed to be of a different nature from those in the latter.
However, the recent re-formulation of the maximum-likelihood (ML) 
decoding problem, for a binary linear code over a discrete memoryless 
channel, as a linear programming problem \cite{FWK} has opened a channel 
through which matroid-theoretic results in combinatorial optimization can be 
applied to coding theory. The key tool in these results is the use
of the decomposition theory of matroids initiated by Seymour \cite{Sey80},
\cite{Sey81}. Based on Seymour's seminal work, Gr\"otschel and Truemper
\cite{GT} showed that the minimization of a linear functional over
the cycle polytope of a binary matroid could be solved in polynomial
time for certain classes of matroids. This immediately implies
that for the corresponding families of codes, the ML decoding problem 
can be solved in time polynomial in the length of the code. Given the 
fact that the ML decoding problem is known to be NP-hard in 
general \cite{BMvT}, the existence of ``non-trivial'' classes of codes
for which ML decoding can be implemented in polynomial time, is
obviously a significant result. However, as we will show in this paper,
for a code family to which the Gr\"otschel-Truemper result applies,
either the dimension or minimum distance of the codes in the family
grows sub-linearly with codelength. Thus, such code families
are not good from a coding-theoretic perspective. However, they do illustrate
the important point that polynomial-time ML decoding is possible. Moreover,
the matroid-theoretic arguments used by Gr\"otschel and Truemper
do not rule out the possibility that there may exist other 
code families for which polynomial-time ML decoding algorithms exist,
which are also good in terms of rate and minimum distance.

The primary goal of this paper is to provide an exposition of the ideas
needed to understand and apply the work of Gr\"otschel and Truemper. 
As mentioned earlier, their work relies upon the machinery provided 
by Seymour's matroid decomposition theory, and so we will first present
that theory in a coding-theoretic setting. Our presentation of this
decomposition theory will be of a tutorial nature. We have attempted
to keep the presentation self-contained to the extent possible;
we do not provide complete proofs of some of the difficult theorems 
that form the basis of the theory.

We provide the relevant definitions and background from matroid theory 
in Section~\ref{matroid_section} of this paper. As explained in that
section, binary matroids and binary linear codes are essentially the same
objects. So, techniques applicable to binary matroids are
directly applicable to binary linear codes as well. In particular,
matroid decomposition techniques can be specialized to codes.

Of central importance in matroid theory is the notion of matroid minors. 
In the context of codes, a minor of a code $\cC$ is any code that can
be obtained from $\cC$ by a sequence of shortening and puncturing 
operations. Minors have received little (if any) attention in coding theory, 
and this seems to be a remarkable oversight 
given the fact that they sometimes capture 
important structural properties of a code. For example, the
presence or absence of certain minors (as stated precisely in 
Theorem~\ref{graphic_code_thm}) decides whether or not a
code is graphic, \emph{i.e.}, has a parity-check matrix that is 
the vertex-edge incidence matrix of some graph. Graphic codes
have been studied previously in the information theory literature
\cite{HB68},\cite{jungnickel}, but the excluded-minor
characterization of these codes appears to have been overlooked
in these studies.

In Section~\ref{conn_section}, we introduce a notion of 
$k$-connectedness for codes, which is again a specialization of 
the corresponding notion for matroids. This is closely related to 
$k$-connectedness in graphs, and interestingly enough, is also 
related to the trellis complexity of a code \cite{For2}.
We do not explore the latter relationship in any detail in this 
paper, instead referring the reader to \cite{kashyap_SIAM},
\cite{kashyap_ITW}, where matroid methods are used to study the
structure of codes with low trellis complexity.

The notion of $k$-connectedness plays an important role in 
Seymour's decomposition theory. An idea of why this is so can 
be obtained from the simple case of 2-connectedness: it follows from
the relevant definitions that a code is not 2-connected if and only if
it is the direct sum of smaller codes. Similar statements can be made
for codes that are not 3- or 4-connected (or more precisely, not
internally 4-connected --- see Definition~\ref{int_4conn_def}) 
via the code-composition operations of 2-sum and 3-sum introduced by 
Seymour \cite{Sey80}. These operations, as well as the $\bar{3}$-sum which
is in a sense the dual operation to the 3-sum, are explained in 
detail in Section~\ref{decomp_section}.

The operations of 2-, 3- and $\bar{3}$-sum have the non-trivial property 
that when two codes $\cC_1$ and $\cC_2$ are composed using one of 
these sums to form a code $\cC$, then $\cC_1$ and $\cC_2$ are 
(up to code equivalence) minors of $\cC$. The relationship between
$k$-connectedness and these sums can then be summarized as follows:
a binary linear code is 2-connected but not 3-connected 
(resp.\ 3-connected, but not internally 4-connected) iff
it can be expressed as the 2-sum (resp.\ 3- or $\bar{3}$-sum) of 
codes $\cC_1$ and $\cC_2$, both of which are equivalent to minors of $\cC$.
It follows immediately from the above facts that 
any binary linear code is either 3-connected and internally 4-connected, 
or can be constructed from 3-connected, internally 4-connected minors 
of it by a sequence of operations of coordinate permutation, 
direct sum, 2-sum, 3-sum and $\bar{3}$-sum. In fact, given any code, 
such a decomposition of the code can be obtained in time polynomial in 
the length of the code.

This code decomposition theory has immediate applications to families
of codes that are minor-closed in the sense that for each code $\cC$ 
in such a family $\mfC$, any code equivalent to a minor of $\cC$ 
is also in $\mfC$. Indeed, a code $\cC$ is in a minor-closed
family $\mfC$ only if the indecomposable (\emph{i.e.}, 3-connected,
internally 4-connected) pieces obtained in the aforementioned 
decomposition of $\cC$ are in $\mfC$.
The above necessary condition is also sufficient if the code family $\mfC$
is additionally closed under the operations of direct sum, 2-sum, 3-sum and 
$\bar{3}$-sum. Thus, membership of an arbitrary code in such a 
family $\mfC$ can be decided in polynomial time iff the membership in 
$\mfC$ of 3-connected, internally 4-connected codes can be decided in 
polynomial time. A formal statement of these and other related facts
can be found in Section~\ref{minor_closed_section} of our paper.

As an illustrative example, we also outline in 
Section~\ref{minor_closed_section} one of the major applications
of Seymour's decomposition theory. This concerns the family of regular
codes which are codes that do not contain as a minor
any code equivalent to the $[7,4]$ Hamming code or its dual. Regular
codes are also characterized by the property that given any
parity-check matrix $H$ of such a code, the 1's in $H$ can be replaced
by $\pm 1$'s in such a way that the resulting $0/\pm 1$ matrix 
is totally unimodular. A totally unimodular matrix is a real matrix all 
of whose square submatrices have determinants in $\{0,1,-1\}$. 
These matrices are of fundamental importance in integer linear 
programming problems \cite{HK}. Seymour \cite{Sey80} proved that a
binary linear code is regular iff it can be decomposed into codes 
that are either graphic, or duals of graphic codes, 
or equivalent to a special $[10,5,4]$ code he called $R_{10}$.

The application of the decomposition theory to linear programming,
and in particular to ML decoding, is the subject of Section~\ref{LP_section}. 
Feldman \emph{et al.}\ showed that the ML decoding problem for a
length-$n$ binary linear code $\cC$ over a discrete memoryless channel 
can be formulated as a minimization problem
$\min \sum_i \gm_i c_i$, where $\gm = (\gm_1,\ldots,\gm_n) \in \R^n$ is a 
certain cost vector derived from the received word and
the channel transition probabilites, and the minimum is taken over
all codewords $\c = (c_1,\ldots,c_n)$ in $\cC$. Now, if $\cC$ is a 
graphic code, then standard graph-theoretic techniques from combinatorial
optimization can be used to find the minimizing codeword in time
polynomial in $n$; a sketch of such an algorithm can be found in 
Appendix~\ref{opt_app}. Gr\"otschel and Truemper \cite{GT} additionally showed 
that this minimization could also be performed in polynomial time for
certain minor-closed code families that are ``almost-graphic'' in a certain
sense. Such a code family $\mfC$ is characterized by the property that 
there exists a \emph{finite} list of codes $\mfD$ such that 
each $\cC \in \mfC$ can be decomposed in polynomial time in such a way
that at each step of the decomposition, one of the pieces is either graphic
or in $\mfD$. 

Gr\"otschel and Truemper gave a polynomial-time algorithm
that takes as input a length-$n$ code $\cC$ from an almost-graphic family 
$\mfC$ and a cost vector $\gm \in \R^n$, and constructs a 
codeword $\c \in \cC$ achieving $\min \sum_i \gm_i c_i$ by
solving related minimization problems over the
pieces of the decomposition that are graphic or in $\mfD$. This algorithm
is also outlined in Appendix~\ref{opt_app}. Thus, the ML decoding
problem can be solved in polynomial time for almost-graphic codes.

Gr\"otschel and Truemper also gave several examples of almost-graphic
families. Interestingly enough, one of these families is that
consisting of codes $\cC$ for which the codeword polytope
(\emph{i.e.}, the convex hull in $\R^n$ of the codewords in the
length-$n$ code $\cC$) is identical to the Koetter-Vontobel 
fundamental polytope \cite{VK05} derived from the entire dual code 
$\cC^\perp$.

Unfortunately, the one truly original result in this paper is a 
negative result. We show that for codes in an almost-graphic 
family, either their dimension or their minimum distance grows 
sub-linearly with codelength. One important implication of this 
is that decoding by linear programming, when applied
to any good error-correcting code, must inevitably hit upon
the occasional pseudocodeword, thus resulting in decoding failure.


We make some concluding remarks in Section~\ref{conclusion}. 
Some of the lengthier or more technical proofs of results
from Sections~\ref{decomp_section} and \ref{LP_section} are 
given in appendices to preserve the flow of the presentation.

\section{Matroids and Codes\label{matroid_section}}
We shall assume familiarity with coding theory; for relevant definitions,
see \cite{sloane}. We will mainly concern ourselves with binary linear
codes, and use standard coding-theoretic notation throughout this paper. 
Thus, an $[n,k]$ code is a code of length $n$ and dimension $k$,
and an $[n,k,d]$ code is an $[n,k]$ code that has minimum distance $d$.
Given a code $\cC$, $\dim(\cC)$ denotes the dimension of $\cC$,
and $\cC^\perp$ denotes the dual code of $\cC$.

The main purpose of this section is to introduce concepts from 
matroid theory that are applicable to coding theory. We will largely follow 
the definitions and notation of Oxley \cite{oxley}. 
We begin with a definition of matroids.
\begin{definition}
A \emph{matroid} $M$ is an ordered pair $(E,\cI)$ consisting of a finite
set $E$ and a collection $\cI$ of subsets of $E$ satisfying the following
three conditions: 
\begin{itemize}
\item[(i)] $\emptyset \in \cI$;
\item[(ii)] if $I \in \cI$ and $J \subset I$, then $J \in \cI$; and 
\item[(iii)] if $I_1,I_2$ are in $\cI$ and $|I_1| < |I_2|$, then there 
exists\footnote{In this paper, we will use $A-B$ to denote the set difference
$A \cap B^c$. The more usual notation $A \setminus B$ 
has been reserved for the matroid operation of ``deletion''.}
$e \in I_2 - I_1$ such that $I_1 \cup \{e\} \in \cI$.
\end{itemize}
\label{matroid_def}
\end{definition}

The set $E$ above is called the \emph{ground set} of the matroid $M$,
and the members of $\cI$ are the \emph{independent sets} of $M$. 
A maximal independent set, \emph{i.e.}, a set $B \in \cI$ such that 
$B \cup \{e\} \notin \cI$ for any $e \in E - B$, is called
a \emph{basis} of $M$. It is a simple consequence of (iii) in 
Definition~\ref{matroid_def} that all bases of $M$ have the same cardinality.
The cardinality of any basis of $M$ is defined to be the \emph{rank} of $M$,
denoted by $r(M)$.

A subset of $E$ that is not in $\cI$ is called a \emph{dependent set}. 
Minimal dependent sets, \emph{i.e.,} dependent sets all of whose
proper subsets are in $\cI$, are called \emph{circuits}. It easily
follows from the definitions that a subset of $E$ is a dependent set
if and only if it contains a circuit.
A dependent set that can be expressed as a disjoint union of 
circuits is called a \emph{cycle}. 

The above definitions of independent and dependent sets, bases and rank
simply try to abstract the notion of independence and dependence, bases
and dimension, respectively, in a vector space over a field. 
Indeed, the most important 
class of matroids for our purposes is the class of binary matroids, 
which are simply vector spaces over the binary field, 
or to put it another way, binary linear codes.

Let $H$ be a binary $m \times n$ matrix, and let 
$\bv_1,\bv_2,\ldots,\bv_n$ denote the column vectors of $H$. 
Set $E = \{1,2,\ldots,n\}$ and take $\cI$ to be the collection of 
subsets $I = \{i_1,i_2,\ldots,i_s\} \subset E$ 
such that the sequence of vectors $\bv_{i_1}, \bv_{i_2}, \ldots, \bv_{i_s}$ 
is linearly independent over the binary field $\F_2$. 
It follows from elementary linear algebra that $(E,\cI)$ 
satisfies the definition of a matroid given above, 
and thus defines a matroid which we shall denote by 
$M[H]$. Note that $r(M[H])$ equals the rank (over $\F_2$) of the matrix $H$.

A matroid $M = (E,\cI)$ is called \emph{binary} 
if it is isomorphic to $M[H]$ for some binary matrix $H$. Here, we
say that two matroids $M = (E,\cI)$ and $M'= (E', \cI')$ are 
\emph{isomorphic}, denoted by $M \cong M'$, if there is a bijection 
$\psi: E \rightarrow E'$ such that for all $J \subset E$, 
it is the case that $J \in \cI$ if and only if $\psi(J) \in \cI'$.

A binary matrix $H$ is also the parity-check matrix of some binary
linear code $\cC$. Note that $r(M[H]) = n - \dim(\cC)$.
The code $\cC$ and the binary matroid $M[H]$ are
very closely related. Recall from coding theory that a codeword 
$\c = (c_1 c_2 \ldots c_n) \in \cC$, $\c \neq \0$, 
is called \emph{minimal} if its support $\supp(\c) = \{i: c_i = 1\}$ 
does not contain as a subset the support of
any other nonzero codeword in $\cC$. It is easily seen that $\c$ 
is a minimal codeword of $\cC$ iff its support 
is a circuit of $M[H]$. It follows from this that
for any $\c \in \{0,1\}^n$, $\c \neq \0$, we have
$\c \in \cC$ iff $\supp(\c)$ is a cycle of $M[H]$.
Furthermore, a routine verification shows that for binary matrices
$H$ and $H'$, $M[H] = M[H']$ iff $H$ and $H'$ are parity-check matrices
of the same code $\cC$. This allows us to associate a unique binary matroid
with each binary linear code $\cC$, and vice versa.

Thus, binary matroids and binary linear codes are essentially the 
same objects. In particular, two codes are 
equivalent\footnote{In coding theory, two binary linear codes 
are defined to be \emph{equivalent} if one can be obtained from 
the other by a permutation of coordinates. In this paper, we will
use the notation $\pi(\cC)$ to denote the code obtained by applying the
coordinate permutation $\pi$ to the code $\cC$.} if and only if their
associated binary matroids are isomorphic. 
This association between codes and binary matroids 
allows us to use tools from matroid theory to study binary linear codes.

While many of the tools used to study matroids have their roots
in linear algebra, there is another source that matroid theory draws from,
namely, graph theory. Indeed, Whitney's founding paper on matroid theory
\cite{whitney} was an attempt to capture the fundamental properties of 
independence that are common to graphs and matrices. 

Let $\cG$ be a finite undirected graph (henceforth simply ``graph'')
with edge set $E$. Define \textsf{cyc} to 
be the collection of edge sets of cycles (\emph{i.e.}, closed walks) in $\cG$. 
Define $I \subset E$ to be independent if $I$ does not contain any 
member of \textsf{cyc} as a subset. Equivalently, $I$ is independent if
the subgraph of $\cG$ induced by $I$ is a forest. Setting $\cI$ to be the
collection of independent subsets of $E$, it turns out that $(E,\cI)$
is a matroid \cite[Proposition~1.1.7]{oxley}. This matroid is 
called the \emph{cycle matroid} of $\cG$, and is denoted by $M(\cG)$.
A matroid that is isomorphic to the cycle matroid of some graph is called
\emph{graphic}.

Clearly, the circuits of $M(\cG)$ are the edge sets of
simple cycles (\emph{i.e.}, closed walks in which no intermediate 
vertex is visited twice) in $\cG$.
The nomenclature ``cycle'' for the disjoint union of circuits in a 
matroid actually stems from its use in the context of graphs.
The bases of $M(\cG)$ are the unions
of edge sets of spanning trees of the connected components
of $\cG$. Hence, $r(M(\cG)) = |V(\cG)| - t$, 
where $V(\cG)$ is the set of vertices of $\cG$, 
and $t$ is the number of connected components of $\cG$.

It is not hard to show that a graphic matroid is binary. Indeed, let $A$ be
the vertex-edge incidence matrix of $\cG$. This is the matrix $[a_{i,j}]$
whose rows and columns are indexed by the vertices and edges, respectively,
of $\cG$, where $a_{i,j} = 1$ if the $j$th edge is incident with the 
$i$th vertex, and $a_{i,j} = 0$ otherwise. It may be verified that
$M(\cG) \cong M[A]$ (see, \emph{e.g.}, \cite[Proposition~5.1.2]{oxley}).

Given a graph $\cG$, we will denote by $\cC(\cG)$ the 
code associated (or identified) with the binary matroid $M(\cG)$.
In other words, $\cC(\cG)$ is the binary linear code that has
the vertex-edge incidence matrix of $\cG$ as 
a parity-check matrix. We will refer to such codes as \emph{graphic codes},
and denote by $\G$ the set of all graphic codes.
Graphic codes have made their appearance previously in the
information theory literature \cite{BH67,HB68} (also see \cite{jungnickel}
and the references therein).

The repetition code of length $n$ is a graphic code; it is the code
obtained from the $n$-cycle $C_n$, the graph consisting of a
single cycle on $n$ vertices.
However, not all binary codes are graphic. For example, it 
can be shown that the [7,4] Hamming code is not graphic. 
It is possible to give a precise characterization of
the codes that are graphic in terms of excluded minors, 
a notion we need to first define. 

There are two well-known ways of obtaining codes of shorter length from
a given parent code. One is via the operation of \emph{puncturing}, in which 
one or more columns are deleted from a generator matrix of the parent code
\cite[p.\ 28]{sloane}. The second method is called 
\emph{shortening}, and involves one or more columns being 
deleted from a parity-check matrix of the parent code \cite[p.\ 29]{sloane}. 
Given a code $\cC$ of length $n$ with generator matrix $G$,
and a subset $J \subset \{1,2,\ldots,n\}$, we will denote by 
$\cC/J$ the code obtained by puncturing the columns of $G$
with indices in $J$, and by $\cC \shorten J$ the code obtained 
by shortening at the columns of $G$ with indices in $J$. 
Note that $\cC/J$ is simply the restriction
of the code $\cC$ onto the coordinates not in $J$, and 
$\cC \shorten J = {(\cC^\perp / J)}^\perp$. 
The notation, though potentially confusing, has been
retained from matroid theory, where the analogues of puncturing and
shortening are called \emph{contraction\/} and \emph{deletion}, respectively.

\begin{definition}
A \emph{minor} of a code $\cC$ is any code obtained from $\cC$ via 
a (possibly empty) sequence of shortening and puncturing operations. 
\label{minor_def}
\end{definition}

It may easily be verified that the precise order in which the 
deletion and puncturing operations are performed is irrelevant. Hence, any
minor of $\cC$ may be unambiguously specified using notation of
the form $\cC/ X \shorten Y$ (or equivalently, $\cC \shorten Y / X$)
for disjoint subsets $X,Y \subset \{1,2,\ldots,n\}$; this notation
indicates that $\cC$ has been punctured at the coordinates indexed by 
$X$ and shortened at the coordinates indexed by $Y$.

The above definition allows a code to be a minor of itself. A minor
of $\cC$ that is not $\cC$ itself is called a \emph{proper minor} of 
$\cC$. Minors have not received much attention in classical coding theory, 
but they play a central role in matroid theory. We will not
touch upon the subject of minors of general matroids, leaving the reader to 
refer to \cite[Chapter~3]{oxley} instead. However, we will briefly
mention how the matroid operations of deletion and contraction specialize
to the cycle matroids of graphs. 

Let $\cG$ be some graph, with edge set $E$. Given $e \in E$, define the graph 
$\cG \delete e$ to be the graph obtained by deleting the edge $e$
along with any vertices that get isolated as a result of deleting $e$.
Also, define $\cG / e$ to be the graph obtained by 
contracting $e$, \emph{i.e.}, deleting $e$ and identifying the two
vertices incident with $e$. The process of obtaining $\cG / e$ 
from $\cG$ is called \emph{edge contraction}, and that of obtaining 
$\cG \shorten e$ from $\cG$ is of course called \emph{edge deletion}.
These operations are inductively extended to define $\cG \delete J$ and 
$\cG / J$ for any $J \subset E$. A minor of a graph $\cG$ is any graph obtained
from $\cG$ via a (possibly empty) sequence of edge deletions and contractions. 

The operations of edge deletion and contraction are
the graphic analogues of code shortening and puncturing, respectively. 
A mathematically precise statement of this is as follows:
given a graph $\cG$ with edge set $E$, 
and any $J \subset E$, we have 
\cite[Equation 3.1.2 and Proposition~3.2.1]{oxley}
\begin{equation*}
\cC(\cG) / J = \cC(\cG / J) \ \ \text{ and } \ \ 
\cC(\cG) \shorten J = \cC(\cG \delete J).
\end{equation*}
It follows that any minor of a graphic code is graphic.

Returning to the question of determining which codes are graphic, the 
answer can be succinctly given in terms of a list of forbidden minors
by the following result of Tutte \cite{Tut59}. 

\begin{theorem}[\cite{Tut59}]
A code is graphic if and only if it does not contain as a minor
any code equivalent to the [7,4] Hamming code or its dual, 
or one of the codes $\cC(K_5)^\perp$ and $\cC(K_{3,3})^\perp$.
\label{graphic_code_thm}
\end{theorem}

In the statement of the above theorem, $K_5$ is the complete
graph on five vertices, while $K_{3,3}$ is the complete bipartite graph
with three vertices on each side. $\cC(K_5)^\perp$ is the $[10,4,4]$
code with generator matrix 
\begin{equation*}
\left[
\begin{array}{cccccccccc}
1 & 0 & 0 & 0 & 1 & 1 & 1 & 0 & 0 & 0 \\
0 & 1 & 0 & 0 & 1 & 0 & 0 & 1 & 1 & 0 \\
0 & 0 & 1 & 0 & 0 & 1 & 0 & 1 & 0 & 1 \\
0 & 0 & 0 & 1 & 0 & 0 & 1 & 0 & 1 & 1 
\end{array}
\right],
\end{equation*}
while $\cC(K_{3,3})^\perp$ is the $[9,5,3]$ code with generator matrix
\begin{equation*}
\left[
\begin{array}{ccccccccc}
1 & 0 & 0 & 0 & 0 & 1 & 1 & 0 & 0 \\
0 & 1 & 0 & 0 & 0 & 0 & 1 & 1 & 0 \\
0 & 0 & 1 & 0 & 0 & 0 & 0 & 1 & 1 \\
0 & 0 & 0 & 1 & 0 & 1 & 0 & 0 & 1 \\
0 & 0 & 0 & 0 & 1 & 1 & 1 & 1 & 1
\end{array}
\right].
\end{equation*}
A proof of the theorem can be found in \cite[Section~13.3]{oxley}. On a
related note, several authors have given algorithms for 
deciding whether or not a given code is graphic \cite{BC80,BW88,Fuj80,Tut60}, 
\cite[Section~10.6]{truemper}. These algorithms run in time polynomial 
in the size of the input, which can be a parity-check matrix for the code.
We will use this fact later in the paper.

\section{Connectedness\label{conn_section}}

As mentioned previously, matroid theory draws upon ideas from graph theory.
A key concept in graph theory is the notion of $k$-connectedness for graphs
\cite[Section~III.2]{bollobas}. 
Given a graph $\cG$, we will let $V(\cG)$ denote
the set of its vertices. A graph is \emph{connected} if any pair of
its vertices can be joined by a path; otherwise, it is \emph{disconnected}.
A maximal connected subgraph of $\cG$ is a \emph{connected component},
or simply \emph{component}, of $\cG$. Let $\cG-W$ denote 
the graph obtained from $\cG$ by deleting the vertices in 
$W$ and all incident edges. If $\cG$ is connected and, for some subset 
$W \subset V(\cG)$, $\cG - W$ is disconnected, then we say that 
$W$ is a \emph{vertex cut} of $\cG$, or that $W$ \emph{separates} $\cG$. 
If $\cG$ is a connected graph that has at least one pair of distinct 
non-adjacent vertices, the \emph{connectivity} $\k(\cG)$ of $\cG$ is defined to
be the smallest integer $j$ for which $\cG$ has a vertex cut $W$ with 
$|W| = j$. If $\cG$ is connected, but has no pair of distinct non-adjacent
vertices, $\k(\cG)$ is defined to be $|V(\cG)| - 1$. Finally, if $\cG$
is disconnected, then we set $\k(\cG) = 0$. For an integer $k > 0$,
$\cG$ is said to be \emph{$k$-connected} if $\k(\cG) \geq k$. Thus, a
graph $\cG$ with $|V(\cG)| \geq 2$ is connected if and only if it is 
1-connected.

The notion of $k$-connectedness of graphs can be extended to matroids,
but it has to be done carefully. One of the problems encountered when
attempting to do so is that 1-connectedness of graphs does not extend
directly to matroids. The reason for this is that for any 
disconnected graph $\cG_1$, there is a connected graph $\cG_2$ such that
$M(\cG_1) \cong M(\cG_2)$ \cite[Proposition~1.2.8]{oxley}. So, the 
link between $k$-connectedness in graphs and that defined below
for matroids begins with the case $k = 2$. 

The definition we present of $k$-connectedness for matroids was formulated
by Tutte \cite{Tut66}. We will once again restrict our attention
to the case of binary matroids (\emph{i.e.}, codes) only. Let 
$\cC$ be a binary linear code of length $n$. We will hereafter use $[n]$
to denote the set of integers $\{1,2,\ldots,n\}$, and for
$J \subset [n]$, we set $J^c = \{i \in [n]: i \notin J\}$.
To further alleviate notational confusion, for $J \subset [n]$, we will define 
$\cC |_J$ to be the restriction of $\cC$ onto its
coordinates indexed by $J$. Equivalently, $\cC |_J = \cC / J^c$, the
latter being the code obtained from $\cC$
by puncturing the coordinates not in $J$.

\begin{definition}
For a positive integer $k$, a partition $(J,J^c)$ of $[n]$ is called
a \emph{$k$-separation of $\cC$} if 
\begin{equation}
\min\{|J|,|J^c|\} \geq k
\label{ksep_eq1}
\end{equation}
and 
\begin{equation}
\dim(\cC|_J) + \dim(\cC|_{J^c}) - \dim(\cC) \leq k-1.
\label{ksep_eq2}
\end{equation}
If $\cC$ has a $k$-separation, then $\cC$ is said to be \emph{$k$-separated}.
\label{ksep_def}
\end{definition}

When equality occurs in (\ref{ksep_eq1}), $(J,J^c)$ is called a 
\emph{minimal $k$-separation}. When equality occurs in (\ref{ksep_eq2}), 
$(J,J^c)$ is called an \emph{exact $k$-separation}. Note that the 
expression on the left-hand side of (\ref{ksep_eq2}) is always
non-negative, since $\dim(\cC |_J) + \dim(\cC |_{J^c}) \geq \dim(\cC)$
for any $J \subset [n]$. This fact easily yields the following result.

\begin{lemma}
$\cC$ is 1-separated iff it is the direct sum of non-empty codes.
\label{1sep_lemma}
\end{lemma}
\begin{proof}
Since $\dim(\cC |_J) + \dim(\cC |_{J^c}) \geq \dim(\cC)$ for any 
$J \subset [n]$, we see from Definition~\ref{ksep_def} that
$(J,J^c)$ is a 1-separation of $\cC$ iff $J,J^c$ are non-empty, 
and $\dim(\cC |_J) + \dim(\cC |_{J^c}) = \dim(\cC)$.
Hence, $(J,J^c)$ is a 1-separation of $\cC$ iff $J,J^c$ are non-empty,
and $\cC$ is the direct sum of $\cC|_J$ and $\cC|_{J^c}$.
\end{proof}

We now give the definition of $k$-connectedness of codes. Note that
this definition starts with $k = 2$.

\begin{definition}
For $k \geq 2$, a code $\cC$ is defined to be \emph{$k$-connected}
if it has no $k'$-separation for any $k' < k$.
\label{conn_def}
\end{definition}

\begin{example}
Let $\cC$ be the [7,3,4] simplex code with generator matrix
$$
G = \left[
\begin{array}{ccccccc}
1 & 0 & 0 & 0 & 1 & 1 & 1 \\
0 & 1 & 0 & 1 & 0 & 1 & 1 \\
0 & 0 & 1 & 1 & 1 & 0 & 1 
\end{array}
\right].
$$
Setting $J = \{1,2,3,7\}$, we see that $(J,J^c)$ forms a 3-separation
of $\cC$. Indeed, the rank of the submatrix of $G$ formed by the 
columns indexed by $J$ is 3, while the rank of the submatrix formed by the 
columns indexed by $J^c$ is 2. In other words, 
$\dim(\cC |_J) = 3$ and $\dim(\cC |_{J^c}) = 2$. 
Thus, both (\ref{ksep_eq1}) and
(\ref{ksep_eq2}) are satisfied with equality, which makes 
$(J,J^c)$ a minimal as well as an exact separation.

It may be verified (for example, by exhaustive search) that there are
no 1- or 2- separations for $\cC$, and all 3-separations are minimal 
and exact. In particular, $\cC$ is 2- and 3-connected, but not 4-connected.
\label{simplex_example}
\end{example}

The quantity, $\dim(\cC|_J) + \dim(\cC|_{J^c}) - \dim(\cC)$, appearing
on the left-hand side of (\ref{ksep_eq2}) in Definition~\ref{ksep_def}
also arises as part of the definition of the state-complexity profile
of a minimal trellis representation of a code 
\cite{For1}--\cite{For3},\cite{horn},\cite{Ksch}.
To be precise, given a length-$n$ code $\cC$, the \emph{state-complexity
profile} of a code \cite[Equation (1)]{horn} is defined to be the vector 
$\s(\cC) = (s_0(\cC),\ldots,s_n(\cC))$, where $s_i(\cC) 
= \dim(\cC|_J) + \dim(\cC|_{J^c}) - \dim(\cC)$ for $J = [i] \subset [n]$.
Here, $[0]$ is defined to be the null set $\emptyset$. It is known 
\cite{For2} that $\s(\cC) = \s(\cC^\perp)$, or equivalently, 
for any $J \subset [n]$,
$$
\dim(\cC|_J) + \dim(\cC|_{J^c}) - \dim(\cC) = 
\dim(\cC^\perp |_J) + \dim(\cC^\perp |_{J^c}) - \dim(\cC^\perp).
$$
As a result, we obtain the interesting and useful fact, 
stated in the proposition below, that $k$-connectedness 
is a property that is invariant under the operation 
of taking code duals. 

\begin{proposition}
Let $\cC$ be a binary linear code of length $n$. 
For any $k \geq 1$, a partition $(J,J^c)$ of $[n]$ is a $k$-separation 
of $\cC$ iff it is a $k$-separation of $\cC^\perp$.
Therefore, for any $k \geq 2$, 
$\cC$ is $k$-connected iff $\cC^\perp$ is $k$-connected.
\label{conn_prop}
\end{proposition}

Consider again the simplex code of length 7 from
Example~\ref{simplex_example}. By Proposition~\ref{conn_prop} above, 
its dual --- the $[7,4]$ Hamming code --- is also
2- and 3-connected, but not 4-connected.

The link between graph and code $k$-connectedness is
strong, but they are not equivalent notions. The closest relation
between the two occurs when $k=2$.
If $\cG$ is a loopless graph without isolated vertices and $|V(\cG)| \geq 3$,
then $\cC(\cG)$ is 2-connected iff $\cG$ is a 2-connected graph 
\cite[Proposition~4.1.8]{oxley}. 
To describe the relation between graph and code connectedness in general,
we define the \emph{connectivity}, $\lambda(\cC)$, of a code $\cC$ 
to be the least positive integer $k$ for which there is a 
$k$-separation of $\cC$, if some $k$-separation exists for $\cC$;  
$\l(\cC)$ is defined to be $\infty$ otherwise. 
Note that $\cC$ is $k$-connected iff $\l(\cC) \geq k$, and 
by Proposition~\ref{conn_prop}, $\l(\cC) = \l(\cC^\perp)$.
It can be shown \cite[Corollary~8.2.7]{oxley}
that for a connected graph $\cG \neq K_3$ 
having at least three vertices,
$$
\l(\cC(\cG)) = \min\{\k(\cG),g(\cG)\},
$$
where $g(\cG)$ denotes the girth (length of shortest cycle) of $\cG$.

Now, our reason for presenting a notion of connectedness for codes
is not just that it extends an idea from graph theory. 
Certain methods of code composition 
have been developed in matroid theory that relate to 2- and 3-separations.
These code composition methods can be considered to be generalizations of 
direct sums, and they allow the result of Lemma~\ref{1sep_lemma} 
to be extended in a non-trivial manner, paving the way for 
the powerful decomposition theory of binary matroids initiated
by Paul Seymour \cite{Sey80}. This decomposition theory allows
one to decompose a binary linear code into smaller codes in 
a reversible manner, in such a way that the smaller codes
are equivalent to minors of the original code. 
As we shall describe in detail in the next section, to find such
a decomposition of a code, we need to find 1-, 2- or 3-separations
in the code, if such separations exist. 

For any fixed positive integer $k$, there are polynomial-time algorithms 
known (see \cite[Section 8.4]{truemper}) that, given a binary linear code 
$\cC$, either find a $k$-separation of $\cC$, 
or conclude that no such separation exists. Here, by ``polynomial-time
algorithm,'' we mean an algorithm that runs in time polynomial 
in the length of $\cC$. For instance, the problem of deciding the 
existence of 1-separations in a code $\cC$ 
is almost trivial. To do so, one takes a matrix $A$ that is either
a generator matrix or a parity-check matrix of $\cC$, 
brings $A$ to reduced row-echelon form (rref), 
removes all-zero rows if they exist, and finally constructs
a certain bipartite graph $BG(A)$. For an $m \times n$ matrix $A$,
the graph $BG(A)$ is defined as follows\footnote{If $A$ is a parity-check
matrix of the code, then $BG(A)$ is simply the corresponding Tanner graph.}:
the vertex set of $BG(A)$ consists of a set of $n$ left vertices 
$\{l_1,\ldots,l_n\}$ and a set of $m$ right vertices 
$\{r_1,\ldots,r_m\}$; an edge connects
the vertices $l_j$ and $r_i$ iff the $(i,j)$th entry of $A$ is 1. 
The code $\cC$ is 2-connected (\emph{i.e.}, 
has no 1-separation) iff $BG(A)$ is connected 
\cite[Lemma~3.3.19]{truemper}.
If $\cC$ is not 2-connected, the connected components of $BG(A)$
induce the required 1-separation.

In general, for fixed integers $k,l$ with $l \geq k$, 
the problem of finding a $k$-separation $(J,J^c)$ of a code, 
with $\min\{|J|,|J^c|\} \geq l$, if it exists,
can be solved in time polynomial in the length of the code, by
an algorithm due to Cunningham and Edmonds (in \cite{cunningham}).
We sketch the idea here. The algorithm is based on the fact that 
the following problem can be solved in time polynomial in $n$:
for codes $\cC_1$ and $\cC_2$ each of length $n$, find 
a partition of $[n]$ that achieves
\begin{equation}
\min\{\dim(\cC_1 |_{J_1}) + \dim(\cC_2 |_{J_2}):\ (J_1,J_2) 
\text{ is a partition of } [n] \}.
\label{mat_int_problem}
\end{equation}
The above problem is solved using the \emph{matroid intersection
algorithm} \cite{E1}, \cite[Section~5.3]{truemper}, 
which we do not describe here. 

The $k$-separation problem of interest to us is equivalent to the
following problem for a fixed integer $l \geq k$: 
given a code $\cC$ of length $n$, find a partition of $[n]$ that achieves 
\begin{equation}
\min\{\dim(\cC |_{J_1}) + \dim(\cC |_{J_2}):\ (J_1,J_2) 
\text{ is a partition of } [n],\ \min\{|J_1|,|J_2|\} \geq l \}.
\label{ksep_min1}
\end{equation}
Indeed, a $k$-separation $(|J|,|J^c|)$ of $\cC$, with 
$\min\{|J|,|J^c|\} \geq l$, exists iff the minimum in (\ref{ksep_min1})
is at most $\dim(\cC) + k-1$. Now, the minimization in (\ref{ksep_min1})
can be solved by finding, for each pair of disjoint $l$-element subsets 
$E_1,E_2 \subset [n]$, the partition $(J_1,J_2)$ of $[n]$ that achieves
\begin{equation}
\min\{\dim(\cC |_{J_1}) + \dim(\cC |_{J_2}):\ (J_1,J_2) 
\text{ is a partition of } [n],\ J_1 \supset E_1,\ J_2 \supset E_2 \}.
\label{ksep_min2}
\end{equation}
If (\ref{ksep_min2}) can be solved in time polynomial in $n$ for each
pair of disjoint $l$-element subsets $E_1,E_2 \subset [n]$, then
(\ref{ksep_min1}) can also be solved in time polynomial in $n$, 
since there are $O(n^{2l})$ pairs $(E_1,E_2)$.

It turns out that (\ref{ksep_min2}) can be solved in time
polynomial in $n$ by converting it to a minimization of the form in 
(\ref{mat_int_problem}), which, as mentioned above, can be solved in 
polynomial time. The trick is to set $\cC_1 = \cC / E_2 \shorten E_1$
and $\cC_2 = \cC /E_1\shorten E_2$, which are both codes
of length $n - |E_1 \cup E_2|$. For notational convenience, we will
let the coordinates of $\cC_1$ and $\cC_2$ retain their
indices from $\cC$, \emph{i.e.}, the set of coordinate indices for
$\cC_1$, as well as for $\cC_2$, is $[n] - (E_1 \cup E_2)$.
It may easily be verified that for $J \subset [n]-(E_1 \cup E_2)$,
$\dim(\cC_i |_J) = \dim(\cC |_{J\cup E_i}) - \dim(\cC|_{E_i})$, $i = 1,2$.
Therefore, for any partition $(J_1,J_2)$ of $[n] - (E_1 \cup E_2)$,
setting $\bar{J_i} = J_i \cup E_i$, $i=1,2$, we have
\begin{equation}
\dim(\cC_1 |_{J_1}) + \dim(\cC_2 |_{J_2}) = 
\dim(\cC |_{\bar{J}_1}) + \dim(\cC |_{\bar{J}_2}) -
\dim(\cC|_{E_1}) - \dim(\cC|_{E_2}),
\label{convert_eq}
\end{equation}
and $(\bar{J_1},\bar{J_2})$ is a partition of $[n]$. Conversely,
for any partition $(\bar{J_1},\bar{J_2})$ of $[n]$, setting
$J_i = \bar{J_i} - E_i$, $i=1,2$, we see that 
$(J_1,J_2)$ forms a partition of $[n] - (E_1 \cup E_2)$, and
(\ref{convert_eq}) is once again satisfied. 

Thus, we see that for a fixed pair of disjoint $l$-element subsets 
$E_1, E_2 \subset [n]$, given a code $\cC$ of length $n$, 
if we set
$\cC_1 = \cC / E_2 \shorten E_1$ and $\cC_2 = \cC /E_1\shorten E_2$, 
then the minimum in (\ref{ksep_min2}) is equal to
$$
\min\{\dim(\cC_1 |_{J_1}) + \dim(\cC_2 |_{J_2}):\ (J_1,J_2) 
\text{ is a partition of } [n] \}
+ \dim(\cC |_{E_1}) + \dim(\cC |_{E_2}).
$$
Since the minima in (\ref{mat_int_problem}) and (\ref{ksep_min2}) just
differ by a constant, the minimization problem (\ref{ksep_min2})
for given $\cC$ and $(E_1,E_2)$ 
is equivalent to the minimization problem (\ref{mat_int_problem})
with $\cC_1 = \cC / E_2 \shorten E_1$ and $\cC_2 = \cC /E_1\shorten E_2$.
Therefore, since (\ref{mat_int_problem}) can be solved in time
polynomial in $n$, so can (\ref{ksep_min2}).

The above sketch does indeed give a polynomial-time algorithm for
determining $k$-separations $(J,J^c)$ with $\min\{|J|,|J^c|\} \geq l$,
but the complexity of the algorithm is $O(n^{2l+\alpha_l})$ for some
constant $\alpha_l$ that arises from the matroid intersection algorithm.
Clearly, this is not very practical even for $l = 3$ or 4, and as we
shall see in the next section, these values of $l$ come up in the
implementation of Seymour's decomposition theory.
A more efficient, albeit more involved, 
algorithm for finding 2- and 3-separations is 
described in \cite[Section~8.4]{truemper}.

As a final remark in this section, we mention that the fact that there
exist algorithms for solving the minimization problems 
(\ref{mat_int_problem})--(\ref{ksep_min2}) that run in time polynomial in $n$
neither contradicts nor sheds any further light on the NP-completeness 
results for the closely related problems considered in \cite{horn}.

\section{Code Composition and Decomposition\label{decomp_section}}

The code composition/decomposition methods described in this section
were developed by Seymour in close analogy with a 
method of composing/decomposing graphs called \emph{clique-sum}. 
In a clique-sum, two graphs, each containing a $K_k$ subgraph 
($k$-clique), are glued together
by first picking a $k$-clique from each graph, sticking the two
cliques together so as to form a single $k$-clique in the composite graph, 
and then deleting some or all of the edges from this clique.
A formal description of clique-sum can be found in 
\cite{Sey80} or in \cite[p.\ 420]{oxley}.

Our exposition of these code composition/decomposition techniques
is based on Seymour's paper \cite{Sey80}. Let $\cC$ and $\cC'$ be
binary linear codes of length $n$ and $n'$, respectively, and let
$m$ be an integer satisfying $0 \leq 2m < \min\{n,n'\}$. We first
define a code $\cC \pl_m \cC'$ as follows: 
if $G = [\g_1 \ \g_2 \ \ldots \ \g_n]$ and 
$G' = [\g'_1 \ \g'_2 \ \ldots \ \g'_{n'}]$ 
are generator matrices of $\cC$ and $\cC'$, respectively,
then $\cC \pl_m \cC'$ is the code with generator matrix
$$
\left[
\begin{array}{ccccccccc}
\g_1 & \ldots & \g_{n-m} & \g_{n-m+1} & \ldots & \g_n & \0 & \ldots & \0 \\
\0 & \ldots & \0 & \g'_1 & \ldots & \g'_m & \g'_{m+1} & \ldots & \g'_{n'}
\end{array}
\right].
$$
Thus, $\cC \pl_m \cC'$ is a binary linear code of length $n+n'-m$. 
This code is almost like a direct sum of $\cC$ and
$\cC'$ except that the two component codes overlap in $m$ positions.
Indeed, when $m = 0$, $\cC \pl_m \cC'$ is the direct sum of $\cC$ and $\cC'$. 

Codewords of $\cC \pl_m \cC'$ are of the form $\c \pl_m \c'$, for
$\c = c_1c_2\ldots,c_n \in \cC$ and $\c' = c'_1c'_2\ldots,c'_{n'} \in \cC'$,
where $\c \pl_m \c' = \hc_1\hc_2\ldots\hc_{n+n'-m}$ is defined to be 
$$
\hc_i = \left\{
\begin{array}{cl}
c_i & \text{for $1 \leq i \leq n-m$} \\
c_i+c'_{i-n+m} & \text{for $n-m+1 \leq i \leq n$} \\
c'_{i-n+m} & \text{for $n+1 \leq i \leq n+n'-m$}
\end{array}
\right.
$$
In other words, $\c \pl_m \c'$ is the binary word of length $n + n' - m$
composed as follows: the first $n-m$ symbols of $\c \pl_m \c'$
are equal to the first $n-m$ symbols of $\c$, the next $m$ symbols of 
$\c \pl_m \c'$ are equal to the coordinatewise modulo-2 sum 
of the last $m$ symbols of $\c$ and the first $m$ symbols of $\c'$, 
and the last $n'-m$ symbols of $\c \pl_m \c'$ 
are equal to the last $n'- m$ symbols of $\c'$.

From $\cC \pl_m \cC'$, we derive a new code, which we temporarily denote
by $\cS_m(\cC,\cC')$, by shortening at the $m$ positions where 
$\cC$ and $\cC'$ are made to overlap. To be precise, let 
$J = \{n-m+1,n-m+2,\ldots,n\}$, and set $\cS_m(\cC,\cC') = 
(\cC \pl_m \cC_2) \shorten J$. Thus, $\cS_m(\cC,\cC')$ is a code
of length $n+n'-2m$ which, by choice of $m$, is greater than $n$ and $n'$.
Once again, note that $\cS_0(\cC,\cC') = \cC \pl_0 \cC' = \cC \oplus \cC'$, 
where $\oplus$ denotes direct sum.

We will actually only be interested in two instances of the above
construction (other than the direct-sum case of $m=0$), one of which is
presented next, and the other is introduced
later in Section~\ref{3sum_section}.

\subsection{2-Sums\label{2sum_section}} The $\cS_m(\cC,\cC')$
construction with $m=1$ is called a 2-sum in certain special cases. 

\begin{definition}
Let $\cC$, $\cC'$ be codes of length at least three, such that
\begin{itemize}
\item[(P1)] $0 \ldots 01$ is not a codeword of $\cC$ or $\cC^\perp$; 
\item[(P2)] $10 \ldots 0$ is not a codeword of $\cC'$ or ${\cC'}^\perp$. 
\end{itemize}
Then, $\cS_1(\cC,\cC')$ is called the \emph{2-sum} of $\cC$ and $\cC'$, 
and is denoted by $\cC \oplus_2 \cC'$.
\label{2sum_def}
\end{definition}

Note that 2-sums are only defined for codes having the
properties (P1) and (P2) listed in Definition~\ref{2sum_def}.
These properties can be equivalently stated as follows:
\begin{itemize}
\item[(P1$'$)] $0 \ldots 0 1$ is not a codeword of $\cC$,
and the last coordinate of $\cC$ is not identically zero; 
\item[(P2$'$)] $1 0 \ldots 0$ is not a codeword of $\cC'$, 
and the first coordinate of $\cC'$ is not identically zero.
\end{itemize}
As we shall see below, (P1$'$) and (P2$'$) are more directly
relevant to an analysis of the 2-sum construction.

\begin{figure}[t]
\centering{\epsfig{file=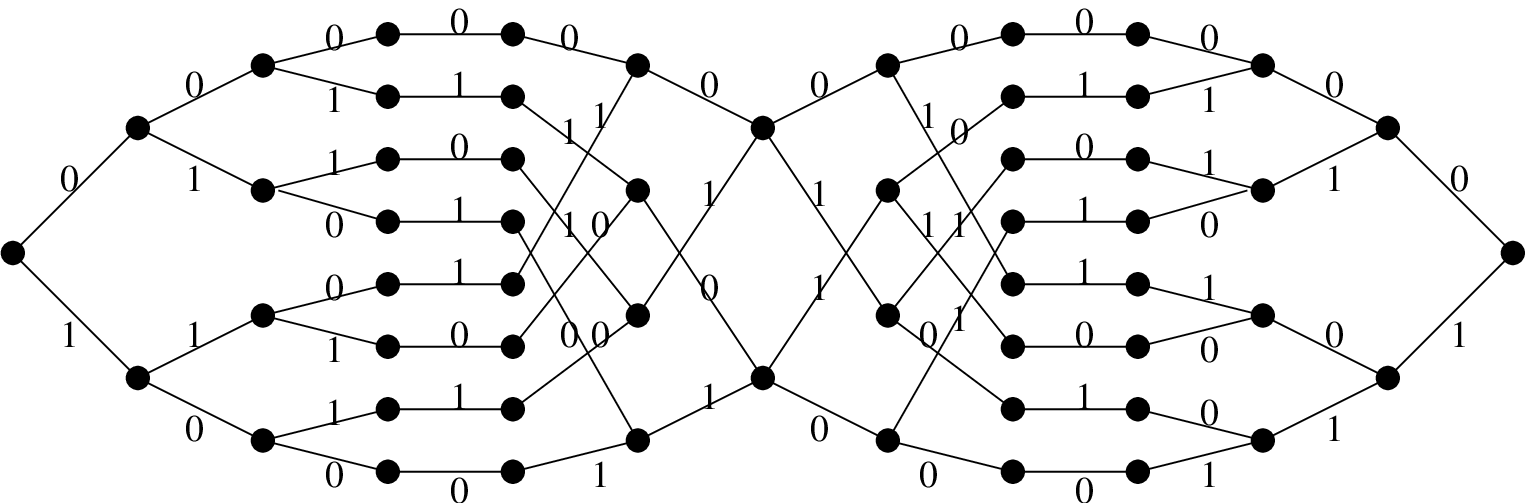, height=3cm}}
\caption{The minimal trellis of the code $\cC \oplus_2 \cC$ of 
Example~\ref{2sum_example}.}
\label{1254_trellis}
\end{figure}

\begin{example}
Let $\cC$ be the $[7,3,4]$ simplex code 
with the generator matrix given in Example~\ref{simplex_example}.
As this code satisfies both (P1) and (P2) in Definition~\ref{2sum_def},
we can define $\cC \oplus_2 \cC$.
Carrying out the 2-sum construction yields
the $[12,5,4]$ code $\cC \oplus_2 \cC$ with generator matrix
$$
\overline{G} = \left[
\begin{array}{cccccccccccc}
1 & 0 & 0 & 0 & 1 & 1 & 0 & 0 & 0 & 1 & 1 & 1 \\
0 & 1 & 0 & 1 & 0 & 1 & 0 & 0 & 0 & 1 & 1 & 1 \\
0 & 0 & 1 & 1 & 1 & 0 & 0 & 0 & 0 & 1 & 1 & 1 \\
0 & 0 & 0 & 0 & 0 & 0 & 1 & 0 & 1 & 0 & 1 & 1 \\
0 & 0 & 0 & 0 & 0 & 0 & 0 & 1 & 1 & 1 & 0 & 1
\end{array}
\right].
$$
The minimal trellis for this code is shown in Figure~\ref{1254_trellis}.
It is easily (from the state-complexity profile of the
minimal trellis, for example) that, with $J = \{1,2,3,4,5,6\}$, 
$(J,J^c)$ is a 2-separation of $\cC \oplus_2 \cC$. 
It may further be verified that $\cC \oplus_2 \cC$ 
has no 1-separation, meaning that it is 2-connected. 
Finally, we note that by shortening
$\cC \oplus_2 \cC$ at the 7th and 8th coordinates, and puncturing
the 9th, 11th and 12th coordinates, we obtain
the simplex code $\cC$ again. In other words,
$\cC$ is a minor of $\cC \oplus_2 \cC$. These observations are not
mere coincidences, as we shall see below.
\label{2sum_example}
\end{example}

The dimension and minimum distance of a 2-sum $\cC \oplus_2 \cC'$ 
can be related to the corresponding parameters of the 
component codes $\cC$ and $\cC'$. Given a binary word $\x$, 
we will use $w(\x)$ to denote its Hamming weight, and for a code $\cC$, 
we will let $d(\cC)$ denote the minimum distance of $\cC$.

\begin{proposition} Let $\cC$ and $\cC'$ be codes for which 
$\cC \oplus_2 \cC'$ can be defined. \\
\emph{(a)} $\dim(\cC \oplus_2 \cC') = \dim(\cC) + \dim(\cC') - 1$. \\
\emph{(b)} $d(\cC \oplus_2 \cC') \leq 
\min\left\{d(\cC \shorten \{n\}),\ d(\cC' \shorten \{1\})\right\}$,
where $n$ is the length of $\cC$. \\
\label{2sum_prop1}
\end{proposition}
\begin{proof} 
Throughout this proof, $n$ and $n'$ denote the lengths of
$\cC$ and $\cC'$, respectively. Also, we shall let $\x \pl \x'$
denote the concatenation of binary words $\x$ and $\x'$.

 (a) By definition, $\cC \oplus_2 \cC' = (\cC \pl_1 \cC') \shorten \{n\}$, 
and so, the 2-sum is isomorphic to the subcode, $\cE$, 
of $\cC \pl_1 \cC'$ consisting of those codewords 
$\hc_1\hc_2 \ldots \hc_{n+n'-1}$ such that $\hc_n = 0$. Since 
the last coordinate of $\cC$ is not identically zero, 
$\cE$ is a proper subcode of $\cC \pl_1 \cC'$, 
and hence, $\dim(\cC \oplus_2 \cC') = \dim(\cE) = \dim(\cC \pl_1 \cC') - 1$.

We claim that the direct sum $\cC \oplus \cC'$ is in fact isomorphic
(as a vector space over $\F_2$) to $\cC \pl_1 \cC'$.
Indeed, consider the map $\phi:\ \cC \oplus \cC' \rightarrow \cC \pl_1 \cC'$
defined via $\phi(\c \pl \c') = \c \pl_1 \c'$, for
$\c \in \cC$ and $\c' \in \cC'$. This 
is a homomorphism onto $\cC \pl_1 \cC'$, but since $0 \ldots 0 1 \notin \cC$
and $10\ldots 0 \notin \cC'$, we have $\ker(\phi) = \{\0\}$, which
shows that $\phi$ is in fact an isomorphism.

Therefore, $\dim(\cC \oplus_2 \cC') = \dim(\cC \pl_1 \cC') - 1
= \dim(\cC \oplus \cC') - 1$, which proves the result.

(b) Let $\hbc$ be a minimum-weight codeword in $\cC \shorten \{n\}$.
Since $\hbc 0 \in \cC$, we have, by construction of $\cC \oplus_2 \cC'$, 
$\hbc \pl_1 \0 \in \cC \oplus_2 \cC'$. Thus, $d(\cC \oplus_2 \cC') 
\leq w(\hbc) = d(\cC \shorten \{n\})$. A similar argument shows that 
$d(\cC \oplus_2 \cC') \leq d(\cC' \shorten \{1\})$.
\end{proof}

An interesting property of 2-sums is that they behave just like
direct sums under the operation of taking code duals. Note that
by virtue of (P1) and (P2) in Definition~\ref{2sum_def}, the 
2-sum of $\cC$ and $\cC'$ can be defined if and only if the 2-sum
of their duals, $\cC^\perp$ and ${\cC'}^\perp$, can be defined. 

\begin{proposition}[\cite{oxley}, Proposition~7.1.20]
Let $\cC$ and $\cC'$ be codes for which 
$\cC \oplus_2 \cC'$ can be defined. Then,
$$
{(\cC \oplus_2 \cC')}^\perp = \cC^\perp \oplus_2 {\cC'}^\perp.
$$
\label{2sum_prop2}
\end{proposition}

This result can be trivially derived from Theorem~7.3 in \cite{For01}, 
but for completeness, we give a simple algebraic proof in 
Appendix~\ref{dual_props_app}. 
As an example, the above result implies that the matrix 
$\overline{G}$ given in Example~\ref{2sum_example} is the
parity-check matrix of the 2-sum of two copies of a [7,4] Hamming code.

While the properties of 2-sums presented above are interesting, the 
usefulness of 2-sums actually stems from the following theorem of Seymour 
\cite{Sey80}, which is a result analogous to Lemma~\ref{1sep_lemma}. 

\begin{theorem}[\cite{Sey80},Theorem~2.6]
If $\cC_1$ and $\cC_2$ are codes of length $n_1$ and
$n_2$, respectively, such that $\cC = \cC_1 \oplus_2 \cC_2$, then 
$(J,J^c)$ is an exact 2-separation of $\cC$, for $J = \{1,2,\ldots,n_1-1\}$.
Furthermore, $\cC_1$ and $\cC_2$ are equivalent to minors of $\cC$.

Conversely, if $(J,J^c)$ is an exact 2-separation of a code $\cC$, then 
there are codes $\cC_1$ and $\cC_2$ of length $|J|+1$ and $|J^c| + 1$,
respectively, such that $\cC$ is equivalent to $\cC_1 \oplus_2 \cC_2$.
\label{2sum_thm}
\end{theorem}

The following corollary is a more concise statement of the above theorem,
and is more in the spirit of Lemma~\ref{1sep_lemma}.

\begin{corollary}
A code $\cC$ has an exact 2-separation 
iff there exist codes $\cC_1$ and $\cC_2$,
both equivalent to proper minors of $\cC$, such that $\cC$ is equivalent
to $\cC_1 \oplus_2 \cC_2$.
\label{2sum_cor1}
\end{corollary}

Another corollary \cite[Theorem~8.3.1]{oxley}, stated next, is a 
consequence of the fact that if $\cC$ is a 2-connected code, 
then any 2-separation of $\cC$
must be exact; if not, the 2-separation would be a 1-separation as well, 
which is impossible as $\cC$ is 2-connected.

\begin{corollary}
A 2-connected code $\cC$ is not 3-connected 
iff there exist codes $\cC_1$ and $\cC_2$,
both equivalent to proper minors of $\cC$, such that $\cC$ is equivalent
to $\cC_1 \oplus_2 \cC_2$.
\label{2sum_cor2}
\end{corollary}

We will not prove Theorem~\ref{2sum_thm} in its entirety,
referring the reader instead to Seymour's original proof, 
or the proof given in Oxley \cite[Section~8.3]{oxley}.
However, we will describe an efficient construction of the components of the 
2-sum when an exact 2-separation $(J,J^c)$ of $\cC$ is given, 
as it is a useful tool in code decomposition. This construction effectively
proves the converse part of the theorem. Our description
is based on the construction given in \cite[Section~8.2]{truemper}.

Let $\cC$ be a code of length $n$ and dimension $k$, specified by
a $k \times n$ generator matrix $G$, and let
$(J,J^c)$ be an exact 2-separation of $\cC$. By permuting coordinates
if necessary, we may assume that $J = \{1,2,\ldots,m\}$ for some $m < n$. 
Let $G|_J$ and $G|_{J^c}$ denote the restrictions of $G$ to the
columns indexed by $J$ and $J^c$, respectively; thus, 
$G = [ G |_J \ \ G|_{J^c}]$. Let $\rank(G|_J) = k_1$ and 
$\rank(G|_{J^c}) = k_2$; since $(J,J^c)$ is an exact 2-separation of $\cC$,
we have $k_1 + k_2 = k + 1$. Bring $G$ into reduced row-echelon form (rref)
over $\F_2$. Permuting coordinates within $J$ and within $J^c$ if necessary, 
$\rref(G)$ may be assumed to be of the form
\begin{equation}
\overline{G} = 
\left[
\begin{array}{cccc}
I_{k_1} & A & \O & B \\
\O & \O & I_{k_2-1} & C
\end{array}
\right],
\label{rref_eq1}
\end{equation}
where $I_j$, for $j = k_1,k_2-1$, denotes the $j \times j$ identity matrix,
$A$ is a $k_1 \times (|J| - k_1)$ matrix, $B$ is a 
$k_1 \times (|J^c| - k_2 + 1)$ matrix, 
$C$ is a $(k_2-1) \times (|J^c| - k_2 + 1)$ 
matrix, and the $\O$'s denote all-zeros matrices of appropriate sizes.
As a concrete example, consider the matrix $\overline{G}$ given in 
Example~\ref{2sum_example}, which is indeed of the above form, with
$|J| = |J^c| = 6$, $k_1 = k_2 = 3$,
$$
A = \left[
\begin{array}{ccc}
0 & 1 & 1 \\ 1 & 0 & 1 \\ 1 & 1 & 0 
\end{array}
\right],\ \ 
B = \left[
\begin{array}{cccc}
0 & 1 & 1 & 1 \\ 0 & 1 & 1 & 1 \\ 0 & 1 & 1 & 1 
\end{array}
\right] \ \ \text{and} \ \ 
C = \left[
\begin{array}{cccc}
1 & 0 & 1 & 1 \\ 1 & 1 & 0 & 1 
\end{array}
\right].
$$

The fact that the submatrix 
$
\left[
\begin{array}{cc}
\O & B \\ I_{k_2-1} & C
\end{array}
\right]
$
must have rank equal to $\rank(G |_{J^c}) = k_2$ implies
that $B$ must have rank 1. Hence, $B$ is actually a matrix with
at least one non-zero row, call it $\b$, and at least one
non-zero column, call it $\tb$. Also, each row of $B$
is either $\0$ or identical to $\b$, and $\tb$
is the length-$k_1$ column vector whose $i$th component
is a 1 if the $i$th row of $B$ is equal to $\b$, and is 0 otherwise.

Now, define 
$$
G_1 = [I_{k_1} \ A \ \ \tb]
$$
and
$$
G_2 = \left[\begin{array}{ccc} 1 & \0^t & \b \\ 
\0 & I_{k_2-1} & C \end{array}\right]
= [I_{k_2} \ C'],
$$
where $\0$ denotes an all-zeros column-vector,
and $C' = \left[\begin{array}{c} \b \\ C \end{array}\right]$.

It is not hard to show that if $\cC_1$ and $\cC_2$ are the codes
generated by $G_1$ and $G_2$, respectively, then $\cC_1 \oplus_2 \cC_2$
is the code generated by the matrix $\overline{G}$ in (\ref{rref_eq1}).
Indeed, carefully going through the construction, it may be verified
that all the rows of $\overline{G}$ are in $\cC_1 \oplus_2 \cC_2$. Hence,
$\dim(\cC_1 \oplus_2 \cC_2) \geq \rank(\overline{G}) = k_1 + k_2 - 1$. 
However, by Proposition~\ref{2sum_prop1}(a), we have that 
$\dim(\cC_1 \oplus_2 \cC_2) = \dim(\cC_1) + \dim(\cC_2) - 1 = 
k_1 + k_2 - 1$. Hence, 
$\dim(\cC_1 \oplus_2 \cC_2) = \rank(\overline{G})$, implying that
$\overline{G}$ must be a generator matrix for $\cC_1 \oplus_2 \cC_2$.

\begin{example}
For the matrix $\overline{G}$ in Example~\ref{2sum_example},
we find the matrices $G_1$ and $G_2$ to be
$$
G_1 = \left[
\begin{array}{ccccccc}
1 & 0 & 0 & 0 & 1 & 1 & 1 \\
0 & 1 & 0 & 1 & 0 & 1 & 1 \\
0 & 0 & 1 & 1 & 1 & 0 & 1
\end{array}
\right],\ \ \ \text{and} \ \ \ 
G_2 = \left[
\begin{array}{ccccccc}
1 & 0 & 0 & 0 & 1 & 1 & 1 \\
0 & 1 & 0 & 1 & 0 & 1 & 1 \\
0 & 0 & 1 & 1 & 1 & 0 & 1
\end{array}
\right], 
$$
which are indeed the generator matrices of the
two simplex codes whose 2-sum is represented by 
$\overline{G}$.
\end{example}
It can also be observed that $\cC_1$ and $\cC_2$ are minors
of $\cC_1 \oplus_2 \cC_2$. Indeed, to obtain $\cC_1$ as a minor of
$\cC_1 \oplus_2 \cC_2$, we proceed as follows. Let $j$ be the 
index of a column of $\overline{G}$ in which the submatrix $B$
has a nonzero column. For the matrix of Example~\ref{2sum_example},
$j$ could be either 10, 11 or 12. Define 
$J_1 = \{|J|+1,|J|+2,\ldots,|J|+k_2-1\}$,
and $J_2 = J^c - (J_1 \cup \{j\})$. Then, 
$\cC_1 = (\cC_1 \oplus_2 \cC_2) \shorten J_1 / J_2$.
To obtain $\cC_2$ as a minor, let $j'$ be the index of a row 
of $\overline{G}$ in which the submatrix $B$ has a nonzero row. 
For the matrix of Example~\ref{2sum_example},
$j'$ could be either 1, 2 or 3. 
The $j'$th column of $\overline{G}$ is of the form 
$[0 \ldots 0\ 1\ 0 \ldots 0]^t$, the single 1 being the $j'$th entry.
Define $J_1' = \{1,2,\ldots,k_1\} - \{j'\}$ and 
$J_2' = \{k_1+1,k_1+2,\ldots,|J|\}$. 
Then, $\cC_2 = (\cC_1 \oplus_2 \cC_2) \shorten J_1' / J_2'$.
Proofs of these statements just involve consistency checking, so 
are left as an easy exercise.

In summary, the procedure described above takes as input a $k \times n$
generator matrix $G$ for $\cC$, and an exact 2-separation $(J,J^c)$ of it,
and produces as output a permutation $\pi$ of the
coordinates of $\cC$, and the generator matrices of
two codes $\cC_1$ and $\cC_2$, such that $\cC = \pi(\cC_1 \oplus_2 \cC_2)$. 
The codes $\cC_1$ and $\cC_2$ are both equivalent to proper minors of $\cC$.
The entire procedure can be carried out in $O(k^2n)$ time, which is the 
run-time complexity of bringing a $k \times n$ matrix to reduced
row-echelon form via elementary row operations. All other parts
of the procedure can be performed in $O(n)$ time; for example,
since $(J,J^c)$ is given, it only takes $O(n)$ time
to find the permutation, $\pi^{-1}$, that takes $\rref(G)$ 
to the matrix $\bar{G}$ in (\ref{rref_eq1}).

A straightforward combination of 
Lemma~\ref{1sep_lemma} and Corollary~\ref{2sum_cor2} yields
the following theorem, which illustrates the utility of the 
matroid-theoretic tools presented so far. 

\begin{theorem}[\cite{oxley}, Corollary~8.3.4]
Every code that is not 3-connected can be constructed from 3-connected
proper minors of it by a sequence of operations of coordinate
permutation, direct sum and 2-sum.
\label{decomp_thm1}
\end{theorem}

The decomposition of a code via direct sums and 2-sums implicit in the above 
theorem can be carried out in time polynomial in the length of the code.
This is due to the following two facts: 
\begin{itemize}
\item[(a)] as described in Section~\ref{conn_section}, 
there are polynomial-time algorithms for finding 1- and 2- separations 
in a code, if they exist; and 
\item[(b)] given an exact 2-separation of a code $\cC$, 
there is a polynomial-time procedure that produces
codes $\cC_1$ and $\cC_2$, both equivalent to proper minors of $\cC$,
and a permutation $\pi$ of the coordinate set of $\cC$,
such that $\cC = \pi(\cC_1 \oplus_2 \cC_2)$.
\end{itemize}

However, direct sums and 2-sums are not enough for our purposes,  
nor were they enough for Seymour's theory of matroid decomposition. 
Seymour also had to extend the graph-theoretic technique of 
3-clique-sum to matroids (in fact, to binary matroids only). The
corresponding operation on binary matroids is called 3-sum.

\subsection{3-Sums\label{3sum_section}}
The special case of the $\cS_3(\cC,\cC')$ construction called 3-sum is
somewhat more complex in definition than the 2-sum.
Recall that for a binary word $\x$, $w(\x)$ denotes its Hamming weight.

\begin{definition}
Let $\cC$, $\cC'$ be codes of length at least seven, such that 
\begin{itemize}
\item[(A1)] no codeword of $\cC$ or ${\cC}^\perp$ is of the form 
$0 \ldots 0\, \x$, where $\x$ is a length-3 word with $w(\x) \in \{1,2\}$; 
\item[(A2)] no codeword of $\cC'$ or ${\cC'}^\perp$ is of the form
$\x\, 0 \ldots 0$, where $\x$ is a length-3 word with $w(\x) \in \{1,2\}$; and
\item[(A3)] $0 \ldots 0 111 \in \cC$ and $111 0 \ldots 0 \in \cC'$.
\end{itemize}
Then, $\cS_3(\cC,\cC')$ is called
the \emph{3-sum} of $\cC$ and $\cC'$, and is denoted by 
$\cC \oplus_3 \cC'$.
\label{3sum_def}
\end{definition}

It is perhaps worth commenting upon the use of the terms ``2-sum'' 
and ``3-sum'' to denote codes of the form $\cS_m(\cC,\cC')$ 
for $m = 1$ and $m=3$, respectively. The nomenclature
stems from the analogy with the $k$-clique-sum of graphs, wherein
two graphs are glued along a $k$-clique. Note that a
2-clique is a single edge (hence $m=1$) and 3-clique is a triangle
of three edges (hence $m=3$). This also explains why we do not
consider an operation of the form $\cS_2(\cC,\cC')$.

3-sums are only defined for codes having the properties (A1)--(A3)
listed in the above definition. It is obvious that an equivalent statement 
of (A1)--(A3) is the following:
\begin{itemize}
\item[(B1)] $0 \ldots 0 111$ is a minimal codeword of $\cC$, 
and $\cC^\perp$ has no nonzero codeword supported entirely within the
last three coordinates of $\cC^\perp$; and
\item[(B2)] $111 0 \ldots 0$ is a minimal codeword of $\cC'$, 
and ${\cC'}^\perp$ has no nonzero codeword supported entirely within the
first three coordinates of ${\cC'}^\perp$.
\end{itemize}
In fact, (B1) and (B2) above
are exact translations of the matroid-theoretic
language used by Seymour in his definition of 3-sum \cite{Sey80}.
Another equivalent way of expressing these conditions is the following:
\begin{itemize}
\item[(B1$'$)] $0 \ldots 0 111$ is a minimal codeword of $\cC$, and the
restriction of $\cC$ onto its last three coordinates is $\{0,1\}^3$; and
\item[(B2$'$)] $111 0 \ldots 0$ is a minimal codeword of $\cC'$, and the
restriction of $\cC'$ onto its first three coordinates is $\{0,1\}^3$.
\end{itemize}
The equivalence of (B1) and (B1$'$) is a consequence of the easily verifiable
fact that if ${0 \ldots 0111} \in \cC$, then 
$\cC^\perp$ has no nonzero codeword supported entirely within the
last three coordinates of $\cC^\perp$ if and only if all possible 3-bit
words appear in the last three coordinates of $\cC$. The equivalence of
(B2) and (B2$'$) is analogous. It follows immediately from 
(B1$'$) and (B2$'$) that $\cC \3s \cC'$ can be defined only if 
$\min\{\dim(\cC),\dim(\cC')\} \geq 3$ and $\max\{d(\cC), d(\cC')\} \leq 3$.

\begin{figure}[t]
\centering{\epsfig{file=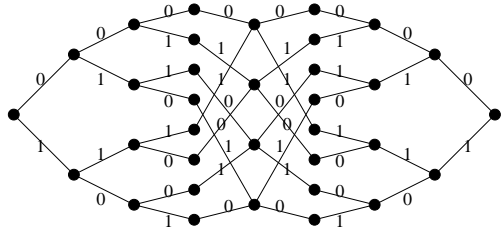, height=3cm}}
\caption{The minimal trellis of the code $\cC \oplus_3 \cC'$ of 
Example~\ref{3sum_example}.}
\label{844_trellis}
\end{figure}

\begin{example}
Let $\cC$ and $\cC'$ be the [7,4] Hamming codes given by the generator
matrices $G$ and $G'$, respectively, below.
$$
G = \left[
\begin{array}{ccccccc}
1 & 1 & 0 & 0 & 0 & 0 & 1 \\
1 & 0 & 1 & 0 & 0 & 1 & 0 \\
0 & 1 & 1 & 0 & 1 & 0 & 0 \\
1 & 1 & 1 & 1 & 0 & 0 & 0 
\end{array}
\right], \ \ \ \ 
G' = \left[
\begin{array}{ccccccc}
1 & 0 & 0 & 0 & 0 & 1 & 1 \\
0 & 1 & 0 & 0 & 1 & 0 & 1 \\
0 & 0 & 1 & 0 & 1 & 1 & 0 \\
0 & 0 & 0 & 1 & 1 & 1 & 1 
\end{array}
\right].
$$
$\cC$ and $\cC'$ satisfy the conditions in Definition~\ref{3sum_def},
so their 3-sum can be defined. The code $\cC \oplus_3 \cC'$ works out to
be the $[8,4,4]$ extended Hamming code with generator matrix
$$
\overline{G} = \left[
\begin{array}{cccccccc}
1 & 0 & 0 & 1 & 0 & 0 & 1 & 1 \\
0 & 1 & 0 & 1 & 0 & 1 & 0 & 1 \\
0 & 0 & 1 & 1 & 0 & 1 & 1 & 0 \\
0 & 0 & 0 & 0 & 1 & 1 & 1 & 1
 \end{array}
\right].
$$
The minimal trellis of this code is shown in Figure~\ref{844_trellis}.

With $J = \{1,2,3,4\}$, $(J,J^c)$ is an exact 
3-separation of $\cC \oplus_3 \cC'$. It may be verified that, in fact, 
$\l(\cC \oplus_3 \cC') = 3$. Furthermore, puncturing any coordinate
of $\cC \oplus_3 \cC'$ yields a $[7,4]$ Hamming code. Thus, $\cC$ and $\cC'$
are (up to code equivalence) minors of $\cC \oplus_3 \cC'$.
\label{3sum_example}
\end{example}

The dimension and minimum distance of $\cC \oplus_3 \cC'$ 
can be related to $\cC$ and $\cC'$ in a manner analogous to
Proposition~\ref{2sum_prop1} for 2-sums.

\begin{proposition}
For codes $\cC$ and $\cC'$ of length $n$ and $n'$ 
for which $\,\cC \3s \cC'$ can be defined, we have
\begin{itemize}
\item[(a)] $\dim(\cC \3s \cC') = \dim(\cC) + \dim(\cC') - 4$.
\item[(b)] $d(\cC \3s \cC') \leq \min\{d(\cC \shorten \{n-2,n-1,n\},
\ d(\cC' \shorten \{1,2,3\})\}$.
\end{itemize}
\label{3sum_prop1}
\end{proposition}
\begin{proof}
We will only prove (a) as the proof of (b) is analogous to the 
proof of Proposition~\ref{2sum_prop1}(b).
Let $\x \pl \x'$ denote the concatenation of binary sequences
$\x$ and $\x'$. 

We prove the proposition by first showing that 
$\dim(\cC \pl_3 \cC') = \dim(\cC) + \dim(\cC') - 1$, and then showing that
$\dim(\cC \oplus_3 \cC') = \dim(\cC \pl_3 \cC') - 3$. The first of these
equalities follows directly from the observation that the mapping
$$
\phi(\c \pl \c') = \c \pl_3 \c',\ \ \ \c \in \cC, \c' \in \cC',
$$
defines a homomorphism from the direct sum
$\cC \oplus \cC'$ onto $\cC \pl_3 \cC'$,
with $\dim(\ker(\phi)) = 1$; indeed, $\ker(\phi)$ consists
of the two words $\0$ and $0 \ldots 0 111 \pl 111 0 \ldots 0$.

To prove that $\dim(\cC \oplus_3 \cC') = \dim(\cC \pl_3 \cC') - 3$, we 
observe that $\cC \oplus_3 \cC'$ is isomorphic to the subcode, $\cE$, of 
$\cC \pl_3 \cC'$ consisting of those codewords 
$\hc_1\hc_2 \ldots \hc_{n+n'-3}$ such that $\hc_{n-2}\hc_{n-1}\hc_n = 000$.
Therefore, $\dim(\cC \oplus_3 \cC') = \dim(\cE)$.

Since the restriction of $\cC$ onto its last three coordinates is $\{0,1\}^3$
(property (B1$'$)), and the restriction of $\cC'$ onto its first three 
coordinates is $\{0,1\}^3$ (property (B2$'$)), the restriction of 
$\cC \pl_3 \cC'$ onto its $(n-2)$th, $(n-1)$th and $n$th coordinates is
also $\{0,1\}^3$. Therefore, $\dim(\cE) = \dim(\cC \pl_3 \cC') - 3$,
and hence, $\dim(\cC \oplus_3 \cC') = \dim(\cC \pl_3 \cC') - 3$, which
completes the proof of the proposition. \end{proof}

An important difference between 2-sums and 3-sums is that the result
of Proposition~\ref{2sum_prop2} does not directly extend to 3-sums. The
reason for this is that for codes $\cC$ and $\cC'$ satisfying (A1)--(A3)
in Definition~\ref{3sum_def}, the 3-sum $\cC^\perp \3s {\cC'}^\perp$ cannot
even be defined. Indeed, while (A1) and (A2) are invariant under the operation
of taking duals, (A3) is not --- if $0 \ldots 0 111 \in \cC$, then
$0 \ldots 0 111 \notin \cC^\perp$. To determine the dual of a 3-sum, we need
to define a ``dual'' operation, namely the $\overline{3}$-sum.

\begin{definition}
Let $\cC$, $\cC'$ be codes of length at least seven, such that 
\begin{itemize}
\item[(A1$'$)] no codeword of $\cC$ or ${\cC}^\perp$ is of the form 
$0 \ldots 0\, \x$, where $\x$ is a length-3 word with $w(\x) \in \{1,2\}$; 
\item[(A2$'$)] no codeword of $\cC'$ or ${\cC'}^\perp$ is of the form
$\x\, 0 \ldots 0$, where $\x$ is a length-3 word with $w(\x) \in \{1,2\}$; and
\item[(A3$'$)] $0 \ldots 0 111 \in \cC^\perp$ and 
$111 0 \ldots 0 \in {\cC'}^\perp$.
\end{itemize}
The \emph{$\bar{3}$-sum} of $\cC$ and $\cC'$, denoted by $\cC \d3s \cC'$,
is defined as 
$$
\cC \d3s \cC' = \bar{\cC}\, \3s\, \bar{\cC'},
$$
where $\bar{\cC} = \cC \, \bigcup \, (0 \ldots 0 111 + \cC)$ and 
$\bar{\cC'} = \cC' \, \bigcup \, (111 0 \ldots 0 + \cC')$.
\label{3barsum_def}
\end{definition}
Note that (A1$'$) and (A2$'$) are identical to (A1) and (A2), respectively.
To ensure that the above definition can in fact be made, it must be verified
that the 3-sum $\bar{\cC}\, \3s\, \bar{\cC'}$ can actually be defined 
for codes $\cC$ and $\cC'$ satisfying (A1$'$)--(A3$'$). So, let $\cC$ and 
$\cC'$ be codes satisfying (A1$'$)--(A3$'$). We need to verify that
$\bar{\cC}$ and $\bar{\cC'}$ satisfy (A1)--(A3) in Definition~\ref{3sum_def}.

By their very definition, $\bar{\cC}$ and $\bar{\cC'}$ satisfy (A3). 
Furthermore, since ${\bar{\cC}}^\perp \subset \cC^\perp$ and 
${\bar{\cC'}}^\perp \subset {\cC'}^\perp$, we see that 
no codeword of ${\bar{\cC}}^\perp$ is of the form $0 \ldots 0\, \x$
as in (A1), and no codeword of ${\bar{\cC'}}^\perp$ is of the form 
$\x\, 0 \ldots 0$ as in (A2). Finally, if $0 \ldots 0\, \x \in \bar{\cC}$
for some length-3 word $\x$ with $w(\x) \in \{1,2\}$, then since
$0 \ldots 0\, \x \notin \cC$ by (A1$'$), it must be that 
$0 \ldots 0\, \x$ is in $0 \ldots 0 111 + \cC$. But in this case, 
$0 \ldots 0\, \bar{\x} \in \cC$, where $\bar{\x} = 111 + \x$. So,
$w(\bar{\x}) \in \{1,2\}$ as well, which is impossible by (A1$'$).
Therefore, $\bar{\cC}$ cannot contain any word of the form 
$0 \ldots 0\, \x$ as in (A1). By analogous reasoning, 
$\bar{\cC'}$ cannot contain any word of the form 
$\x\,0 \ldots 0$ as in (A2). We have thus verified that
$\bar{\cC}$ and $\bar{\cC'}$ satisfy (A1)--(A3), and so
$\bar{\cC}\, \3s\, \bar{\cC'}$ can be defined. 

Note that (A3$'$) implies that $0 \ldots 0 111 \notin \cC$ and 
$111 0 \ldots 0 \notin \cC'$, and hence, 
$\dim(\bar{\cC}) = \dim(\cC) + 1$ and $\dim(\bar{\cC'}) = \dim(\cC') + 1$.
Therefore, by virtue of Proposition~\ref{3sum_prop1}, we have the
following result.

\begin{lemma}
$\dim(\cC \d3s \cC') = \dim(\cC) + \dim(\cC') - 2$.
\label{3barsum_lemma}
\end{lemma}

The $\bar{3}$-sum is the dual operation to 3-sum, in a sense made 
precise by the proposition below, the proof of which we defer to
Appendix~\ref{dual_props_app}. This result is stated, without
explicit proof, in \cite[p.\ 316]{GT} and \cite[p.\ 184]{truemper}; 
Truemper \cite{GT,truemper} refers to 3-sum and $\bar{3}$-sum as 
$\Delta$-sum and $Y$-sum, respectively. The result can also be derived
from Theorem~7.3 in \cite{For01}.

\begin{proposition}
Let $\cC$ and $\cC'$ be codes for which 
$\cC \oplus_3 \cC'$ can be defined. Then, 
$$
{(\cC \3s \cC')}^\perp = \cC^\perp \d3s {\cC'}^\perp.
$$
\label{3sum_prop2}
\end{proposition}

It follows from the last result that a code that is 
expressible as a 3-sum can also be expressed as a
$\bar{3}$-sum. Indeed, if $\cC = \cC_1 \oplus_3 \cC_2$, then 
by the above proposition, $\cC^\perp = \cC_1^\perp \d3s \cC_2^\perp$.
The latter, by definition, is $\bar{\cC_1^\perp} \3s \bar{\cC_2^\perp}$,
and so again taking duals and using the above proposition, we obtain
$$
\cC = {\left(\bar{\cC_1^\perp} \3s \bar{\cC_2^\perp}\right)}^\perp
 = \left(\bar{\cC_1^\perp}\right)^\perp \d3s 
\left(\bar{\cC_2^\perp}\right)^\perp.
$$
We record this as a corollary to Proposition~\ref{3sum_prop2}.

\begin{corollary}
If $\cC_1 \3s \cC_2$ can be defined, then
$$
\cC_1 \3s \cC_2 = \left(\bar{\cC_1^\perp}\right)^\perp \d3s 
\left(\bar{\cC_2^\perp}\right)^\perp.
$$
\label{3sum_cor}
\end{corollary}

Having presented some of the simpler properties of 3-sums, we next state
a highly non-trivial result of Seymour that illustrates how 3-sums
are to be used. The statement of this result is the 3-sum analogue of 
Theorem~\ref{2sum_thm}, but there are some important differences between 
the two that we will point out after stating the result.

\begin{theorem}[\cite{Sey80},Theorems~2.9 and 4.1]
If $\cC_1$ and $\cC_2$ are codes of length $n_1$ and
$n_2$, respectively, such that $\cC = \cC_1 \oplus_3 \cC_2$, then 
$(J,J^c)$ is an exact 3-separation of $\cC$ 
for $J = \{1,2,\ldots,n_1-3\}$. Furthermore, if $\cC$ is 3-connected, 
then $\cC_1$ and $\cC_2$ are equivalent to proper minors of $\cC$. 

Conversely, if $(J,J^c)$ is an exact 3-separation of a code $\cC$, with
$\min\{|J|,|J^c|\} \geq 4$, then there are codes 
$\cC_1$ and $\cC_2$ of length $|J|+3$ and $|J^c| + 3$,
respectively, such that $\cC$ is equivalent to $\cC_1 \oplus_3 \cC_2$.
\label{3sum_thm}
\end{theorem}

A couple of key differences between the statements of 
Theorems~\ref{2sum_thm} and \ref{3sum_thm} must be stressed. 
For a code to be expressible as a 2-sum, it is sufficient that 
there exist an exact 2-separation.
However, to make the analogous conclusion about 3-sums, 
Theorem~\ref{3sum_thm} not only asks for the existence of an 
exact 3-separation $(J,J^c)$, but also adds the additional hypothesis
that $\min\{|J|,|J^c|\} \geq 4$. We will have more
to say about this a little later. 

There is a second major difference between the statements of the 
two theorems. Theorem~\ref{2sum_thm} states that if 
$\cC = \cC_1 \oplus_2 \cC_2$, then $\cC_1$ and $\cC_2$ are 
always minors of $\cC$, up to coordinate permutation. However, 
when $\cC = \cC_1 \oplus_3 \cC_2$, Theorem~\ref{3sum_thm} 
imposes the condition that $\cC$ be 3-connected
in order to conclude that $\cC_1$ and $\cC_2$ are equivalent to 
minors of $\cC$. If the 3-connectedness requirement for $\cC$ is dropped,
the conclusion does not hold in general, as the following example shows.

\begin{figure}[t]
\centering{\epsfig{file=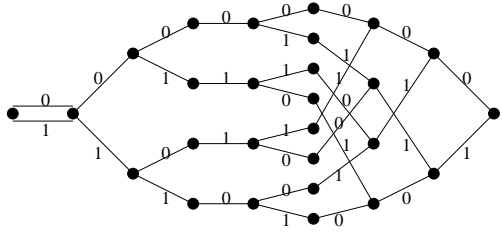, height=3cm}}
\caption{The minimal trellis of the code $\cC = \cC_1 \oplus_3 \cC_2$ of 
Example~\ref{not_3conn_example}.}
\label{841_trellis}
\end{figure}

\begin{example}
Take $\cC_1$ to be the $[7,4,1]$ code with generator matrix
$$
G_1 = \left[
\begin{array}{ccccccc}
1 & 0 & 0 & 0 & 0 & 0 & 0 \\
0 & 1 & 0 & 1 & 0 & 0 & 1 \\
0 & 0 & 1 & 1 & 0 & 1 & 0 \\
1 & 1 & 1 & 0 & 1 & 0 & 0 
\end{array}
\right],
$$
and let $\cC_2$ be the $[7,4,3]$ Hamming code with generator matrix
$$
G_2 = \left[
\begin{array}{ccccccc}
1 & 0 & 0 & 0 & 1 & 1 & 0 \\
0 & 1 & 0 & 0 & 1 & 0 & 1 \\
0 & 0 & 1 & 0 & 0 & 1 & 1 \\
0 & 0 & 0 & 1 & 1 & 1 & 1 
\end{array}
\right].
$$
These codes satisfy (A1)--(A3) of Definition~\ref{3sum_def}, and
their 3-sum, $\cC_1 \oplus_3 \cC_2$,
is the $[8,4,1]$ code $\cC$ generated by 
$$
G = \left[
\begin{array}{cccccccc}
1 & 0 & 0 & 0 & 0 & 0 & 0 & 0 \\
0 & 1 & 0 & 1 & 0 & 0 & 1 & 1 \\
0 & 0 & 1 & 1 & 0 & 1 & 0 & 1 \\
0 & 0 & 0 & 0 & 1 & 1 & 1 & 1
\end{array}
\right].
$$

The minimal trellis of this code is shown in Figure~\ref{841_trellis}.
Note that $\cC$ is not 3-connected. In fact, it is not 
even 2-connected --- it has a 1-separation $(J,J^c)$ with $J = \{1\}$. 
Now, $\cC_1$ can be obtained as a minor of $\cC$
by puncturing $\cC$ at the last coordinate. However, $\cC_2$ is not a 
minor of $\cC$, since puncturing $\cC$ at any coordinate does not
yield a $[7,4,3]$ code, and shortening always yields 
a code of dimension less than 4.
\label{not_3conn_example}
\end{example}

We mention in passing that if $\cC = \cC_1 \oplus_3 \cC_2$, 
then the fact that $\cC_1$ and $\cC_2$ are equivalent to minors of $\cC$ 
whenever $\cC$ is 3-connected is far more difficult to prove 
(see \cite[Section 4]{Sey80}) than the corresponding part of 
Theorem~\ref{2sum_thm}.

We observed above that the mere existence
of an exact 3-separation is not enough for Theorem~\ref{3sum_thm} 
to conclude that a code $\cC$ is expressible as a 3-sum;
the 3-separation $(J,J^c)$ must also satisfy 
$\min\{|J|,|J^c|\} \geq 4$, \emph{i.e.}, must
not be minimal\footnote{The definition of a minimal 
$k$-separation is given immediately after Definition~\ref{ksep_def}.}. 
It is also implicit in the statement of Theorem~\ref{3sum_thm} 
that the existence of a non-minimal 3-separation 
is a necessary condition for a code to be a 3-sum.
Indeed, if $\cC = \cC_1 \oplus_3 \cC_2$, then $\cC_1$ and $\cC_2$ must each
have length at least 7, as per Definition~\ref{3sum_def}. So, with
$J = \{1,2,\ldots,n_1-3\}$ as in the statement of Theorem~\ref{3sum_thm},
it must be true that $|J| \geq 4$, and similarly, $|J^c| \geq 4$.

The following definition allows for a compact statement of a corollary to
Theorem~\ref{3sum_thm} along the lines of Corollary~\ref{2sum_cor2}.

\begin{definition}
A 3-connected code is \emph{internally 4-connected} if 
all its 3-separations are minimal.
\label{int_4conn_def}
\end{definition}

Internal 4-connectedness is a notion that lies properly between 
3-connectedness and 4-connectedness --- a 3-connected code that is 
not 4-connected can be internally 4-connected. Note that any 
3-separation in a 3-connected code must be exact, and so we can state
the following corollary to Theorem~\ref{3sum_thm}.

\begin{corollary}
A 3-connected code $\cC$ is not internally 4-connected 
iff there exist codes $\cC_1$ and $\cC_2$,
both equivalent to proper minors of $\cC$, such that $\cC$ is equivalent
to $\cC_1 \oplus_3 \cC_2$.
\label{3sum_cor1}
\end{corollary}

As in the case of 2-sums, we provide an efficient construction of the 
components of the 3-sum when an exact, non-minimal 
3-separation $(J,J^c)$ of $\cC$ is given. This construction 
furnishes a proof of the converse part of Theorem~\ref{3sum_thm}. 
Our description is based loosely on the constructions given 
in \cite[Section~8.3]{truemper} and \cite{GT}.

Let $\cC$ be a code of length $n$ and dimension $k$, specified by
a $k \times n$ generator matrix $G$, and let
$(J,J^c)$ be an exact 3-separation of $\cC$, with $|J| \geq 4$
and $|J^c| \geq 4$. By permuting coordinates
if necessary, we may assume that $J = \{1,2,\ldots,m\}$ for some $m$
such that $4 \leq m \leq n-4$. 
Let $G|_J$ and $G|_{J^c}$ denote the restrictions of $G$ to 
the columns indexed by $J$ and $J^c$, respectively;
thus, $G = [ G |_J \ \ G|_{J^c}]$. Let $\rank(G|_J) = k_1$ and 
$\rank(G|_{J^c}) = k_2$; since $(J,J^c)$ is an exact 3-separation of $\cC$,
we have $k_1 + k_2 = k + 2$. Note that $k_1 \leq k$ implies that 
$k + k_2 \geq k_1 + k_2 = k + 2$, so that $k_2 \geq 2$; similarly,
$k_1 \geq 2$.

Bring $G$ into reduced row-echelon form (rref)
over $\F_2$. Permuting coordinates within $J$ and within $J^c$ if necessary, 
$\rref(G)$ may be assumed to be of the form
\begin{equation}
\overline{G} = 
\left[
\begin{array}{cccc}
I_{k_1} & A & \O & B \\
\O & \O & I_{k_2-2} & C
\end{array}
\right],
\label{rref_eq2}
\end{equation}
where $I_j$, for $j = k_1,k_2-2$, denotes the $j \times j$ identity matrix,
$A$ is a $k_1 \times (|J| - k_1)$ matrix, $B$ is a 
$k_1 \times (|J^c| - k_2 + 2)$ matrix, 
$C$ is a $(k_2-2) \times (|J^c| - k_2 + 2)$ 
matrix, and the $\O$'s denote all-zeros matrices of appropriate sizes.
As a concrete example, consider the matrix $\overline{G}$ given in 
Example~\ref{3sum_example}, which is indeed of the above form, with
$|J| = |J^c| = 4$, $k_1 = k_2 = 3$,
$$
A = \left[
\begin{array}{c}
1 \\ 1 \\ 1
\end{array}
\right],\ \ 
B = \left[
\begin{array}{ccc}
0 & 1 & 1 \\ 1 & 0 & 1 \\ 1 & 1 & 0
\end{array}
\right] \ \ \text{and} \ \ 
C = \left[
\begin{array}{ccc}
1 & 1 & 1 
\end{array}
\right].
$$

The fact that the submatrix 
$
\left[
\begin{array}{cc}
\O & B \\ I_{k_2-2} & C
\end{array}
\right]
$
must have rank equal to $\rank(G |_{J^c}) = k_2$ implies
that $B$ must have rank 2. Hence, $B$ has two
linearly independent rows, call them $\x$ and $\y$, which form
a basis of the row-space of $B$. In particular,
each row of $B$ is either $\0$, $\x$, $\y$ or $\x+\y$.

Now, define the $(k_1+1) \times (|J|+3)$ matrix
$$
G_1 = \left[\begin{array}{ccc}
I_{k_1} & A & D \\
\0 & \0 & \1
\end{array}
\right],
$$
where $D$ is a $k_1 \times 3$ matrix whose $i$th row is defined as
$$
\text{$i$th row of $D$} = 
\left\{
\begin{array}{cl}
000 & \text{if $i$th row of $B$ is $\0$} \\
001 & \text{if $i$th row of $B$ is $\x$} \\
010 & \text{if $i$th row of $B$ is $\y$} \\
100 & \text{if $i$th row of $B$ is $\x + \y$};
\end{array}
\right.
$$
and the bottom row of $G_1$, represented by $[\0\ \ \0\ \ \1]$,
is simply $0 \ldots 0 1 1 1$.

Next, define the $(k_2+1) \times (|J^c|+3)$ matrix
$$
G_2 = \left[\begin{array}{ccc} I_3 & \O & X \\ 
\O & I_{k_2-2} & C \end{array}\right]
= [I_{k_2+1} \ C''],
$$
where 
$$
X = \left[\begin{array}{c}
\x+\y \\ \y \\ \x
\end{array}\right] 
\ \ \ \text{and} \ \ \ 
C'' = \left[\begin{array}{c} X \\ C \end{array}\right].
$$

A straightforward verification yields that if 
$\cC_1$ and $\cC_2$ are the codes generated by $G_1$ and $G_2$, 
respectively, then $\dim(\cC_1) = k_1+1$,
$\dim(\cC_2) = k_2 + 1$, and $\cC_1,\cC_2$ satisfy properties (A1)--(A3)
in Definition~\ref{3sum_def}, so $\cC_1 \oplus_3 \cC_2$ can be defined.
The construction of $G_1$ and $G_2$ above is carefully crafted to ensure that
all the rows of $\overline{G}$ are in $\cC_1 \oplus_3 \cC_2$. 
Hence, $\dim(\cC_1 \oplus_3 \cC_2) \geq \rank(\overline{G}) = k_1 + k_2 - 2$. 
However, by Proposition~\ref{3sum_prop1}, we have that 
$\dim(\cC_1 \oplus_3 \cC_2) = \dim(\cC_1) + \dim(\cC_2) - 4 = 
k_1 + k_2 - 2$. Hence, 
$\dim(\cC_1 \oplus_3 \cC_2) = \rank(\overline{G})$, implying that
$\overline{G}$ must be a generator matrix for $\cC_1 \oplus_3 \cC_2$.
Note that according to Theorem~\ref{3sum_thm}, if $\cC$ is 3-connected,
then $\cC_1$ and $\cC_2$ are equivalent to proper minors of $\cC$.

The procedure described above can be formalized into an algorithm that 
takes as input a $k \times n$ generator matrix $G$ for $\cC$, 
and an exact, non-minimal 3-separation $(J,J^c)$ of it,
and produces as output a permutation $\pi$ of the
coordinates of $\cC$, and the generator matrices of
two codes $\cC_1$ and $\cC_2$, such that $\cC = \pi(\cC_1 \oplus_3 \cC_2)$. 
The codes $\cC_1$ and $\cC_2$ are both equivalent to proper minors of $\cC$.
This procedure can be carried out
in $O(k^2n)$ time for the same reasons as in the 2-sum case.

For the matrix $\overline{G}$ in Example~\ref{3sum_example},
we find the matrices $G_1$ and $G_2$ to be
$$
G_1 = \left[
\begin{array}{ccccccc}
1 & 0 & 0 & 1 & 1 & 0 & 0 \\
0 & 1 & 0 & 1 & 0 & 1 & 0 \\
0 & 0 & 1 & 1 & 0 & 0 & 1 \\
1 & 1 & 1 & 1 & 0 & 0 & 0 
\end{array}
\right],\ \ \ \text{and} \ \ \ 
G_2 = \left[
\begin{array}{ccccccc}
1 & 0 & 0 & 0 & 0 & 1 & 1 \\
0 & 1 & 0 & 0 & 1 & 0 & 1 \\
0 & 0 & 1 & 0 & 1 & 1 & 0 \\
0 & 0 & 0 & 1 & 1 & 1 & 1
\end{array}
\right], 
$$
which are indeed generator matrices of the
two Hamming codes whose 3-sum is represented by 
$\overline{G}$.

It must be pointed out that Theorem~\ref{3sum_thm} also holds when
the 3-sums in its statement are replaced by $\bar{3}$-sums. This is
a consequence of Proposition~\ref{conn_prop}, which shows that $(J,J^c)$
is a 3-separation of a code $\cC$ iff it is a 3-separation of the
dual code $\cC^\perp$. Hence, applying Theorem~\ref{3sum_thm} to $\cC^\perp$,
and dualizing via Proposition~\ref{3sum_prop2}, we see that the 
3-sums in the statement of Theorem~\ref{3sum_thm} can be 
replaced by $\bar{3}$-sums. In particular, we also have the 
following corollary to Theorem~\ref{3sum_thm}.

\begin{corollary}
A 3-connected code $\cC$ is not internally 4-connected 
iff there exist codes $\cC_1$ and $\cC_2$,
both equivalent to proper minors of $\cC$, such that $\cC$ is equivalent
to $\cC_1 \d3s \cC_2$.
\label{3sum_cor2}
\end{corollary}

Putting together Theorem~\ref{decomp_thm1} with
\ref{3sum_cor1} and \ref{3sum_cor2}, we obtain the 
following theorem, which summarizes the code decomposition
theory presented up to this point. 

\begin{theorem}
A binary linear code either is 3-connected and internally 4-connected,
or can be constructed from 3-connected, internally 4-connected
proper minors of it by a sequence of operations of coordinate
permutation, direct sum, 2-sum and 3-sum (or $\bar{3}$-sum).
\label{decomp_thm2}
\end{theorem}

The decomposition of Theorem~\ref{decomp_thm2}
can be carried out in time polynomial in the length of the code, since
\begin{itemize}
\item[(a)] 
the decomposition of Theorem~\ref{decomp_thm1} can be carried out in 
polynomial time; 
\item[(b)] as mentioned in Section~\ref{conn_section},
there are polynomial-time algorithms for finding non-minimal 
3-separations (\emph{i.e.}, 3-separations $(J,J^c)$ with 
$\min\{|J|,|J^c|\} \geq 4$) in a code, if they exist; and
\item[(c)] there is a polynomial-time procedure that, given an exact,
non-minimal 3-separation of a 3-connected code $\cC$, produces codes 
$\cC_1$ and $\cC_2$, both equivalent to proper minors of $\cC$,
and a permutation $\pi$ of the coordinate set of $\cC$,
such that $\cC = \pi(\cC_1 \oplus_3 \cC_2)$ or, via Corollary~\ref{3sum_cor},
$\cC = \pi(\cC_1 \d3s \cC_2)$.
\end{itemize}
Note that the theorem does not guarantee uniqueness of 
the code decomposition.

\subsection{Code-Decomposition Trees}
A binary tree is a convenient data structure for storing a
decomposition of a code via direct sums, 2-sums, and 3-sums.
Recall that a proper (or full) binary tree is a rooted tree such 
that every node of the tree has either zero or two children.
We will drop the adjective ``proper'' as proper binary trees
are the only kind of binary trees we are interested in. 
A node without any children is called a \emph{leaf}. Each 
non-leaf node has two children, and we will distinguish between
the two, calling one the \emph{left} child and the other 
the \emph{right} child.


Let $\cC$ be a binary linear code. A \emph{code-decomposition tree}
for $\cC$ is a binary tree $\cT$ defined as follows.
Each node \sfv of $\cT$ stores a triple 
(\textsf{v.code},\textsf{v.perm},\textsf{v.sum}), where
\textsf{v.code} is a binary linear code, \textsf{v.perm} is either
NULL or a permutation of the coordinate set of \textsf{v.code}, and
\textsf{v.sum} $\in \{\odot,\oplus,\oplus_2,\oplus_3,\d3s\}$. 
For each node \sfv of $\cT$, the triple 
(\textsf{v.code},\textsf{v.perm},\textsf{v.sum})
must adhere to the following rules:
\begin{itemize}
\item[(R1)] if \sfv is the root node, then \textsf{v.code} is the code
$\cC$ itself;
\item[(R2)] \textsf{v.perm} $=$ NULL iff \sfv is a leaf;
\item[(R3)] \textsf{v.sum} $= \odot$ iff \sfv is a leaf;
\item[(R4)] if \sfv is a non-leaf node, then 
\begin{itemize}
\item[(i)] \textsf{lchild.code} and 
\textsf{rchild.code} are proper minors of \textsf{v.code}; and
\item[(ii)] the permutation \textsf{v.perm} applied to the sum
$$
(\text{\sf lchild.code})\ \
\text{\sf v.sum}\ \
(\text{\sf rchild.code})
$$
yields \textsf{v.code},
\end{itemize}
where \textsf{lchild} and \textsf{rchild} above respectively refer to 
the left and right children of \textsf{v}.
\end{itemize}
In particular, (R4) ensures that for any node \textsf{v}
other than the root node in the tree, \textsf{v.code} 
is a proper minor of $\cC$.


We will identify the leaves of any code-decomposition tree 
with the codes that they store. Theorem~\ref{decomp_thm2} 
guarantees that each code $\cC$ has a code-decomposition tree 
in which each leaf is 3-connected and internally 4-connected. Such
a code-decomposition tree will be called \emph{complete}.
A complete code-decomposition tree for $\cC$ can be constructed
in time polynomial in the length of the code $\cC$, since
the decomposition of Theorem~\ref{decomp_thm2}
can be carried out in polynomial time. Note that if $\cC$
itself is 3-connected and internally 4-connected, then there is
exactly one code-decomposition tree for it, which is the tree
consisting of the single node $\cC$ --- or more precisely,
the single node that stores $(\cC,\text{NULL},\odot)$.

Finally, a code-decomposition tree is called \emph{3-homogeneous}
(resp.\ \emph{$\bar{3}$-homogeneous}) if for each non-leaf node
\sfv in the tree, if \textsf{v.sum} $\in \{\oplus,\2s,\3s\}$
(resp.\ \textsf{v.sum} $\in \{\oplus,\2s,\d3s\}$).
Thus, in a 3-homogeneous (resp.\ $\bar{3}$-homogeneous) tree, 
no $\bar{3}$-sums (resp.\ 3-sums) are used. It follows from
Corollaries~\ref{3sum_cor1} and \ref{3sum_cor2} that if
a code-decomposition tree has a node of the form \sfv $= (\cC,\pi,\d3s)$,
with $\cC$ being 3-connected, then the subtree with \sfv as a 
root can be replaced with a different subtree in which the root node is 
\textsf{v'} $= (\cC,\pi',\3s)$. Therefore, every code has
a complete, 3-homogeneous (or $\bar{3}$-homogeneous) code-decomposition
tree, which again can be constructed in time polynomial in the
length of the code.


Having described in detail Seymour's decomposition theory in the context
of binary linear codes, we now turn to some of its applications.
This theory mainly derives its applications from families of codes
that are minor-closed, and such families form the subject of the next section.

\section{Minor-Closed Families of Codes\label{minor_closed_section}}

A family $\mfC$ of binary linear codes is defined to be
\emph{minor-closed} if for each $\cC \in \mfC$, every code equivalent
to a minor of $\cC$ is also in $\mfC$. Note that this definition
automatically implies that a minor-closed family, $\mfC$, of codes is
closed under code equivalence, \emph{i.e.}, if $\cC \in \mfC$, then 
all codes equivalent to $\cC$ are also in $\mfC$. 

A non-trivial example of a minor-closed family is the set of all 
graphic codes, since any minor of a graphic code is graphic.
We will encounter other examples of minor-closed families 
(regular codes and geometrically perfect codes)
further on in this paper. We mention in passing another 
interesting example of such a family --- codes of bounded
trellis state-complexity. Recall from Section~\ref{conn_section}
that the state-complexity profile of a length-$n$ code $\cC$ 
is defined to be the vector $\s(\cC) = (s_0(\cC),\ldots,s_n(\cC))$, 
where $s_i(\cC) = \dim(\cC|_J) + \dim(\cC|_{J^c}) - \dim(\cC)$ 
for $J = [i] \subset [n]$. Define $s_{\max}(\cC) = 
\max_{i \in [n]} s_i(\cC)$. For a fixed integer $w > 0$, let $TC_w$ denote
the family of codes $\cC$ such that there exists a code $\cC'$ equivalent
to $\cC$ with $s_{\max}(\cC') \leq w$. Then, $TC_w$ is minor-closed
\cite{kashyap_SIAM}, \cite{kashyap_ITW}. A similar statement holds
for the family of codes that have a cycle-free normal realization 
(cf.\ \cite{For03}) whose state-complexity is bounded by $w$.

A general construction of minor-closed families is obtained
by fixing a collection $\cF$ of codes, and defining
$\mfC_\cF$ to be the set of all codes $\cC$ such that no minor
of $\cC$ is equivalent to any $\cC' \in \cF$. As an example, let 
$\cF = \{\cH_7,\cH_7^\perp,\cC(K_5)^\perp,\cC(K_{3,3})^\perp\}$,
where $\cH_7$ is the $[7,4]$ Hamming code\footnote{From now on,
$\cH_7$ will always denote the $[7,4]$ Hamming code.}.
By Theorem~\ref{graphic_code_thm}, $\mfC_\cF$ in this 
case is precisely the family of graphic codes. 
It is clear that $\mfC_\cF$ is a minor-closed family for any fixed $\cF$. 
In fact, every minor-closed family can be obtained in this manner. 
Indeed, let $\mfC$ be a minor-closed family of codes. 
A code $\mfD$ is said to be an \emph{excluded minor} of 
$\mfC$ if $\cD \notin \mfC$, but every proper minor of $\cD$
is in $\mfC$. It is not hard to verify that a code $\cC$ 
is in $\mfC$ iff no minor of $\cC$ is an excluded minor of $\mfC$. 
Theorem~\ref{graphic_code_thm} is an example of such an 
\emph{excluded-minor characterization}, and we will see 
more such examples (Theorems~\ref{regular_thm1} and \ref{P_Q_thm})
further below. Thus, taking $\cF$ to be the collection of 
all excluded minors of $\mfC$, we have that $\mfC = \mfC_{\cF}$.
A tantalizing conjecture of Robertson and Seymour asserts that any 
minor-closed family $\mfC$ of binary linear codes
has only \emph{finitely many} excluded minors \cite[Conjecture~1.2]{GGW_survey}. 

Let $\mfC$ be a minor-closed family of codes. By Theorem~\ref{decomp_thm2}, 
every $\cC \in \mfC$ can be constructed
from 3-connected, internally 4-connected codes in $\mfC$ using
direct sums, 2-sums, and 3- or $\bar{3}$-sums. 
The converse need not always be true, \emph{i.e.}, it is not
necessarily true that if a code $\cC$ has a decomposition via
direct sums, 2-sums, and 3- or $\bar{3}$-sums into codes in $\mfC$,
then $\cC \in \mfC$. Of course, the converse does hold if $\mfC$
is also closed under the operations of direct sum, 2-sum, 3-sum
and $\bar{3}$-sum. As usual, $\mfC$ is defined to be closed
under direct sum (resp.\ 2-sum, 3-sum, $\bar{3}$-sum) if 
for any pair of codes in $\cC$, their direct sum 
(resp.\ 2-sum, 3-sum, $\bar{3}$-sum, if it can be defined) 
is also in $\mfC$. We summarize this in the following proposition.

\begin{proposition}
Let $\mfC$ be a minor-closed family of codes that is also
closed under the operations of direct sum, 2-sum, 3-sum and
$\bar{3}$-sum. Then, the following are equivalent for a code $\cC$.
\begin{itemize}
\item[(i)] $\cC$ is in $\mfC$.
\item[(ii)] The leaves of some code-decomposition tree 
for $\cC$ are in $\mfC$.
\item[(iii)] The leaves of some complete, 3-homogeneous 
or $\bar{3}$-homogeneous code-decomposition tree for $\cC$ 
are in $\mfC$.
\item[(iv)] The leaves of every code-decomposition tree 
for $\cC$ are in $\mfC$.
\end{itemize}
\label{minor_closed_prop}
\end{proposition}

\begin{proof}
(i) implies (iv) since the leaves of any code-decomposition tree 
of $\cC$ are minors of $\cC$, and $\mfC$ is minor-closed.
The implications (iv) $\Rightarrow$ (iii) and (iii) $\Rightarrow$ (ii)
are trivial. (ii) implies (i) since $\mfC$ is closed under 
direct-sums, 2-sums, 3-sums and $\bar{3}$-sums.
\end{proof}

Since a complete code-decomposition tree of any code $\cC$ can be constructed
in time polynomial in the length of $\cC$, we have the following corollary
to the above result.

\begin{corollary}
Let $\mfC$ be a minor-closed family of codes that is also
closed under the operations of direct sum, 2-sum, 3-sum
and $\bar{3}$-sum. Then the following are equivalent statements.
\begin{itemize}
\item[(i)] It can be decided in polynomial time whether
or not a given code $\cC$ is in $\mfC$.
\item[(ii)] It can be decided in polynomial time whether
or not a given 3-connected, internally 4-connected code $\cC$ 
is in $\mfC$.
\label{minor_closed_cor}
\end{itemize}
\end{corollary}

The first major application of results such as the above ---
the application which was in fact the motivation for Seymour's 
matroid decomposition theory --- relates to totally unimodular matrices. 
A real matrix $A$ is said to be \emph{totally unimodular}
if the determinant of every square submatrix of $A$ is in $\{0,1,-1\}$.
In particular, each entry of a totally unimodular matrix is in 
$\{0,1,-1\}$. Such matrices are of fundamental importance in
combinatorial optimization and network flow problems, because 
total unimodularity is closely related to integer linear programming
\cite{HK}.

A binary matrix is defined to be \emph{regular} if its 1's can
be replaced by $\pm 1$'s in such a way that the resulting
matrix is totally unimodular. Consequently, a binary linear code
is defined to be \emph{regular} if it has a regular parity-check matrix.
It turns out that for a regular code, \emph{every} parity-check matrix
is regular \cite[Corollary~9.2.11]{truemper}. Furthermore, given
a regular binary matrix $B$, there is a polynomial-time algorithm 
that converts $B$ to a totally unimodular matrix by assigning
signs to the 1's in $B$ \cite[Corollary~9.2.7]{truemper}. 
Thus, regular codes form the key to understanding total unimodularity.
The following theorem, due to Tutte \cite{Tut58}, provides an elegant
excluded-minor characterization of regular codes.

\begin{theorem}
A binary linear code is regular iff it does not contain as a minor
any code equivalent to the [7,4] Hamming code or its dual.
\label{regular_thm1}
\end{theorem}

It follows from the theorem that the family of regular codes,
which we will denote by $\mfR$, is
minor-closed, since it is of the form $\mfC_\cF$ for 
$\cF = \{\cH_7,\cH_7^\perp\}$. Furthermore, $\mfR$ is
closed under the taking of code duals, \emph{i.e.}, the dual of a
regular code is also regular. This is because a code $\cC$ contains
$\cH_7$ as a minor iff its dual $\cC^\perp$ contains $\cH_7^\perp$
as a minor. It can further be shown \cite[p.\ 437]{oxley}
that $\mfR$ is closed under the operations of direct sum,
2-sum, 3-sum and $\bar{3}$-sum.

Note that by Theorem~\ref{graphic_code_thm},
$\mfR$ contains the family of graphic codes, and hence, the family of
\emph{co-graphic} codes as well, which are codes whose duals are graphic.
Using a long and difficult argument, Seymour \cite{Sey80} proved that 
the 3-connected, internally 4-connected codes in $\mfR$ are either graphic,
co-graphic, or equivalent to a particular isodual
code that he called $R_{10}$, which is neither graphic nor co-graphic. 

\begin{theorem}[\cite{oxley}, Corollary~13.2.6]
If $\cC$ is a 3-connected, internally 4-connected regular code, then
$\cC$ is either graphic, co-graphic, or equivalent to $R_{10}$, which
is the $[10,5,4]$ code with parity-check matrix
$$
\left[
\begin{array}{cccccccccc}
1 & 1 & 0 & 0 & 1 & 1 & 0 & 0 & 0 & 0 \\
1 & 1 & 1 & 0 & 0 & 0 & 1 & 0 & 0 & 0 \\
0 & 1 & 1 & 1 & 0 & 0 & 0 & 1 & 0 & 0 \\
0 & 0 & 1 & 1 & 1 & 0 & 0 & 0 & 1 & 0 \\
1 & 0 & 0 & 1 & 1 & 0 & 0 & 0 & 0 & 1 
\end{array}
\right].
$$
\label{regular_thm2}
\end{theorem}

Thus, Seymour's decomposition theory shows that any regular code
(and so by assignment of signs, any totally unimodular matrix)
can be constructed by piecing together --- via direct sums, 2-sums, 
and 3-sums or $\bar{3}$-sums --- graphic codes, co-graphic
codes, and codes equivalent to $R_{10}$. Also, membership
in the family of regular codes can be decided in polynomial time.
Indeed, as mentioned at the end of Section~\ref{matroid_section}, there are 
polynomial-time algorithms for deciding whether or not a given code 
is graphic. Given an $m \times n$ parity-check matrix $H$ for a 
code, a generator matrix for the code can be computed using elementary
row operations on $H$ in $O(m^2n)$ time. Thus, the dual of a code
can be determined in polynomial time, and hence it can be decided
in polynomial time whether or not a given code is co-graphic.
Hence, from Corollary~\ref{minor_closed_cor} and 
Theorem~\ref{regular_thm2}, it follows that there is a polynomial-time
algorithm for determining whether or not a given code is regular.
The best such algorithm known is due to Truemper \cite{T90},
which runs in $O((m+n)^3)$ time. Truemper's algorithm is also based
on Seymour's decomposition theory, but it implements a highly 
efficient procedure for carrying out the decomposition.

While the application of Seymour's decomposition theory to regular codes
is interesting, it is not very useful, perhaps, from a coding-theoretic 
perspective. However, in the next section, we give an application that 
should be of some interest to a coding theorist.

\section{Application: ML Decoding\label{LP_section}}

The recent work of Feldman, Wainwright and Karger \cite{FWK} shows 
that ML decoding of a binary linear code $\cC$
over a discrete memoryless channel can be formulated as a
linear program (LP). Recall that the ML decoding problem
is: given a received word $\y$ at the channel output,
find a codeword $\x\in\cC$ that maximizes the probability, 
$\Pr[\y|\x]$, of receiving $\y$ conditioned on the event
that $\x$ was transmitted. 
As observed by Feldman \emph{et al.}, under the assumption
of a discrete memoryless channel, given a received word 
$\y=y_1 y_2 \ldots y_n$, the
problem of determining ${\arg\max}_{\x\in\cC} \Pr[\y|\x]$ is
equivalent to the problem of finding 
${\arg\min}_{\x\in\cC} \la\mathbf{\gm},\x\ra$,
where $\mathbf{\gm} = (\gm_1,\gm_2,\ldots,\gm_n)$ is given by
\begin{equation}
\gm_i = \log \left(\frac{\Pr[y_i|x_i=0]}{\Pr[y_i|x_i=1]}\right)
\label{gamma_def}
\end{equation}
and $\la\cdot\, , \cdot\ra$ is the standard inner product on $\R^n$.
Here, for the inner product $\la\mathbf{\gm},\x\ra$ to make sense,
a binary codeword $\x=x_1 x_2 \ldots x_n \in\cC$ is 
identified with the real vector $(x_1,x_2,\ldots,x_n) \in \{0,1\}^n
\subset \R^n$. 

The above formulation shows ML decoding to be 
equivalent to the minimization of a linear function
over a finite set $\cC \subset \{0,1\}^n$. Let $P(\cC)$
be the \emph{codeword polytope} of $\cC$, \emph{i.e.},
the convex hull in $\R^n$ of the finite set $\cC$.
It can be shown that the set of vertices of $P(\cC)$ coincides with $\cC$.
The key point now is that over a polytope $P$, a linear function 
$\phi$ attains its minimum value $\phi_{\min} = \min \{\phi(\x): \x \in P\}$
at a vertex of $P$. In particular,
$
{\min}_{\x\in \cC} \la\mathbf{\gm},\x\ra
= {\min}_{\x\in P(\cC)} \la\mathbf{\gm},\x\ra,
$
Thus, ML decoding is equivalent to finding a vertex of the polytope
$P(\cC)$ that achieves ${\min}_{\x\in P(\cC)} \la\mathbf{\gm},\x\ra$,
which is a classic LP.

However, ML decoding of an arbitrary code is known to be NP-hard
\cite{BMvT}. So, in general, solving the above LP over the codeword polytope
is also NP-hard. A strategy often followed in such a situation
is to ``relax'' the problem. The idea is to 
look for a polytope that contains the code as a subset of its vertex set,
but which has some property that allows an LP defined over it 
to be solved more easily. Such a polytope is called a \emph{relaxation}
of the codeword polytope $P(\cC)$.

A certain relaxation of the codeword polytope has received much
recent attention \cite{FWK},\cite{VK05},\cite{ST06}. This is
the polytope which, given a code $\cC$ of length $n$, 
and a subset $H \subset \cC^\perp$, is defined as
$$
Q(H) = \bigcap_{\h \in H} P(\h^\perp),
$$
where $P(\h^\perp)$ is the codeword polytope of the code 
$\h^\perp = \{\c \in \F_2^n:\ \la \h,\c \ra \equiv 0 \pmod 2\}$.
Note that since $\cC \subset \h^\perp$ for any $\h \in \cC^\perp$,
we have that $P(\cC) \subset \bigcap_{\h \in H} P(\h^\perp) = Q(H)$ 
for any $H \subset \cC^\perp$.
In particular, $P(\cC) \subset Q(\cC^\perp)$ for any code $\cC$.

For any $H \subset \cC^\perp$, the polytope $Q(H)$ contains $\cC$
as a subset of its vertex set, $\cV(H)$. This is because 
$\cC \subset Q(H) \cap \{0,1\}^n$, and since $Q(H)$ is 
contained within the $n$-cube $[0,1]^n$, we also have 
$Q(H) \cap \{0,1\}^n \subset \cV(H)$. Thus, $Q(H)$ is indeed 
a relaxation of $P(\cC)$. Consequently the LP 
$\min_{\x \in Q(H)} \la \gm, \x\ra$, where $\gm$ is the vector defined
via (\ref{gamma_def}), constitutes a relaxation of the LP 
that represents ML decoding.

Now, any standard LP-solving algorithm requires that the LP 
to be solved have its constraints be represented via 
linear inequalities. The advantage of using
the relaxation $Q(H)$ is that there is a convenient such 
representation of the constraint $\x \in Q(H)$. 
The polytope $Q(H)$ can also be expressed as 
(see \emph{e.g.}\ \cite[Theorem~4]{FWK} or \cite[Lemma~26]{VK05}),
\begin{equation}
Q(H) = \bigcap_{\h \in H} \Pi(\supp(\h)),
\label{Q_eq1}
\end{equation}
where for $S \subset [n]$, $\Pi(S)$ denotes the polyhedron
\begin{equation}
\Pi(S) = \bigcap_{\stackrel{J \subset S}{\mbox{\tiny $|J|$ odd}}}
\left\{(x_1,\ldots, x_n) \in [0,1]^n:\ \sum_{j \in J} x_j - 
\sum_{i \in S \setminus J} x_i
\leq |J| - 1\right\}.
\label{Q_eq2}
\end{equation}
The efficiency of a practical LP solver (like, say, the simplex or ellipsoid
algorithm) depends on the size of the LP representation, which is
proportional to the number of variables and linear inequalities forming
the constraints. If $H$ above consists of the rows of a parity-check
matrix of a low-density parity-check (LDPC) code, then the representation
of $Q(H)$ given by (\ref{Q_eq1})--(\ref{Q_eq2}) has size linear in 
the codelength $n$. So, the ellipsoid algorithm, for example, 
would be guaranteed to solve the LP 
$\min_{\x \in Q(H)} \la \gm, \x\ra$ in time polynomial in $n$.
However, as we explain next, this LP is no longer equivalent to
ML decoding in general.

Let $\cJ(H) = \cV(H) \cap \{0,1\}^n$ 
denote the set of \emph{integral vertices} (\emph{i.e.}, vertices all of 
whose coordinates are integers) of $Q(H)$. We noted above that 
$\cC \subset \cJ(H)$. If $H$ is a spanning subset (over $\F_2$) 
of $\cC^\perp$, so that the vectors in $H$ form (the rows of) 
a parity-check matrix of $\cC$, then we in fact have $\cC = \cJ(H)$. 
This is because if $\x \in \{0,1\}^n$ is not in $\cC$, 
then $\x \notin \h^\perp$ for some $\h \in H$, 
and hence, $\x \notin P(\h^\perp) \supset Q(H)$.
The polytope $Q(H)$ in this case is the ``fundamental polytope''
of Vontobel and Koetter \cite{VK05} (or equivalently, 
the ``projected polytope'' $\overline{Q}$ 
of Feldman \emph{et al}.\ \cite[p.\ 958]{FWK}). The fact 
that $\cC = \cJ(H)$ for such a polytope $Q(H)$ implies that 
the polytope has the following ``ML certificate'' 
property \cite[Proposition~2]{FWK}: if the LP 
$\min_{\x \in Q(H)} \la \gm, \x\ra$, where $\gm$ is the vector defined
via (\ref{gamma_def}), attains its minimum at some $\x \in \cJ(H)$,
then $\x$ is guaranteed to be the ML codeword. However, it is 
possible that the above LP attains its minimum at some 
non-integral vertex $\x \in \cV(H) - \cJ(H)$, in which case decoding 
via linear programming over $Q(H)$ fails. The non-integral vertices 
of $Q(H)$ are called ``pseudocodewords''.

It is naturally of interest to know when a code $\cC$ has 
a fundamental polytope $Q(H)$ (for some spanning subset $H$ of 
$\cC^\perp$) without pseudocodewords. For such codes, ML
decoding can be exactly implemented as an LP over $Q(H)$. Clearly,
$Q(H)$ has no pseudocodewords iff $\cC = \cV(H)$, or equivalently, 
$P(\cC) = Q(H)$. But since $P(\cC) \subset Q(\cC^\perp) \subset Q(H)$,
this obviously implies that we must have $P(\cC) = Q(\cC^\perp)$.
Conversely, if $P(\cC) = Q(\cC^\perp)$, then we may simply take 
$H = \cC^\perp$ to obtain a fundamental polytope $Q(H)$ without 
pseudocodewords. We record this observation as a lemma.

\begin{lemma}
Let $\cC$ be a binary linear code. There exists a 
spanning subset, $H$, of $\cC^\perp$ such that the polytope
$Q(H)$ has no pseudocodewords iff $P(\cC) = Q(\cC^\perp)$.
\label{pseudocodeword_lemma}
\end{lemma}

A code $\cC$ for which $P(\cC) = Q(\cC^\perp)$ holds will be called
\emph{geometrically perfect}, and we will denote by 
$\mfG$ the family of all such codes. So the question then is:
which codes are geometrically perfect? 
An answer to this was provided by Barahona and Gr\"otschel \cite{BG}, 
who showed that the relationship $P(\cC) = Q(\cC^\perp)$ is equivalent
to Seymour's ``sums-of-circuits'' property for binary matroids
\cite{Sey81}. The following theorem is thus equivalent to 
Seymour's characterization of binary matroids with the sums-of-circuits 
property.

\begin{theorem}[\cite{BG}, Theorem~3.5] 
A binary linear code $\cC$ is geometrically perfect iff 
$\cC$ does not contain as a minor any code equivalent to 
$\cH_7^\perp$, $R_{10}$ or $\cC(K_5)^\perp$.
\label{P_Q_thm}
\end{theorem}

By the above theorem, $\mfG$
is minor-closed. Moreover, since none of $\cH_7^\perp$, $R_{10}$ or 
$\cC(K_5)^\perp$ is graphic, no graphic code can contain any of
them as a minor, and so graphic codes are geometrically perfect.

\begin{figure}[t]
\centerline{\epsfig{file=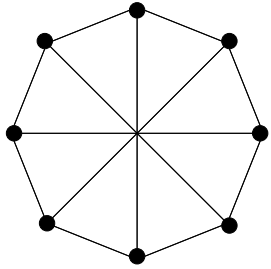, width=3.5cm}}
\caption{The graph $V_8$.}
\label{V8_fig}
\end{figure}

Gr\"otschel and Truemper \cite[Section~4]{GT} showed that $\mfG$ is closed
under the operations of direct sum, 2-sum and $\bar{3}$-sum,
but is not closed under 3-sum. They also observed \cite[p.\ 326]{GT}
that any code in $\mfG$ can be constructed via direct sums,
2-sums and $\bar{3}$-sums from graphic codes and copies
of the codes $\cH_7$, $\cC(K_{3,3})^\perp$ and $\cC(V_8)^\perp$, where
$V_8$ is the graph in Figure~\ref{V8_fig}. Indeed, this result
is implied by Theorems~6.4, 6.9 and 6.10 in \cite{Sey81}. It is not
hard to verify that the codes $\cH_7$, $\cC(K_{3,3})^\perp$ and 
$\cC(V_8)^\perp$ are in fact in $\mfG$. Putting these facts together,
we obtain the following theorem.

\begin{theorem}
For a binary linear code $\cC$, the following are equivalent statements.
\begin{itemize}
\item[(i)] $\cC$ is geometrically perfect, \emph{i.e.},
$P(\cC) = Q(\cC^\perp)$.
\item[(ii)] Each leaf in some complete, $\bar{3}$-homogeneous 
code-decomposition tree for $\cC$ is either graphic, or equivalent
to one of the codes $\cH_7$, $\cC(K_{3,3})^\perp$ and $\cC(V_8)^\perp$.
\item[(iii)] Each leaf in every complete, $\bar{3}$-homogeneous 
code-decomposition tree for $\cC$ is either graphic, or equivalent
to one of the codes $\cH_7$, $\cC(K_{3,3})^\perp$ and $\cC(V_8)^\perp$.
\end{itemize}
\label{mfG_decomp_thm}
\end{theorem}

\begin{proof}
(i) implies (iii) follows directly from the fact that
any code in $\mfG$ can be constructed via direct sums,
2-sums and $\bar{3}$-sums from graphic codes and copies (up 
to equivalence) of the codes $\cH_7$, $\cC(K_{3,3})^\perp$ 
and $\cC(V_8)^\perp$. The implication 
(iii) $\Rightarrow$ (ii) is trivial. Finally,
(ii) $\Rightarrow$ (i) holds since graphic codes and 
the codes $\cH_7$, $\cC(K_{3,3})^\perp$ and $\cC(V_8)^\perp$ are
all in $\mfG$, and $\mfG$ is closed under direct sum, 2-sum 
and $\bar{3}$-sum.
\end{proof}

Since a complete, $\bar{3}$-homogeneous code-decomposition tree for a code
can be constructed in polynomial time, and testing for 
graphicness or equivalence to 
$\cH_7$, $\cC(K_{3,3})^\perp$ and $\cC(V_8)^\perp$ can also be
carried out in polynomial time, we have the following corollary to
the above theorem.

\begin{corollary}
It can be decided in polynomial time whether or not a given code
$\cC$ is geometrically perfect, \emph{i.e.}, has the property 
$P(\cC) = Q(\cC^\perp)$.
\label{mfG_cor}
\end{corollary}

However, this is only half the story. If $\cC$ is a geometrically
perfect code, the algorithm guaranteed by the above result will determine
this to be the case, but will not produce a ``small'' subset 
$H \subset \cC^\perp$ such that $P(\cC) = Q(H)$. The only information 
we would have is that $H$ can be taken to be the \emph{entire} dual code
$\cC^\perp$. While it would then be true that the LP 
$\min_{\x \in Q(\cC^\perp)} \la\gm,\x\ra$ is equivalent to ML decoding,
no known LP-solving algorithm could be guaranteed to efficiently solve 
this LP. This is because the representation of $Q(\cC^\perp)$ 
given by (\ref{Q_eq1})--(\ref{Q_eq2}) has size exponential in 
the codelength $n$. Fortunately, as we shall describe next, for the 
family, $\mfG$, of geometrically perfect codes, ML decoding can always be 
implemented in time polynomial in codelength, \emph{not} using
an LP-solving algorithm, but by means of a combinatorial optimization
algorithm that uses code decompositions. \\[-6pt]

In a series of papers \cite{T85}--\cite{T90} (see also \cite{truemper}),
Truemper carried out a careful examination of matroid decompositions,
from which a particularly interesting observation concerning 
geometrically perfect codes could be inferred. 
Let $\mfG_0$ be the sub-family of $\mfG$ 
that consists of all graphic codes and codes equivalent to one of 
$\cH_7$, $\cC(K_{3,3})^\perp$ and $\cC(V_8)^\perp$.
As observed in the proof of Corollary~6.6 in \cite{GT}, Truemper's analysis of 
matroid decompositions could be used to show that a 2-connected 
code $\cC \in \mfG - \mfG_0$ is equivalent to 
either $\cC_1 \oplus_2 \cC_2$ or $\cC_1 \d3s \cC_2$, for some
$\cC_1 \in \mfG_0$ and some 2-connected code $\cC_2 \in \mfG$,
both of which are equivalent to minors of $\cC$. 
The decomposition of $\cC$ into $\cC_1$ and $\cC_2$, along with the
coordinate permutation that takes $\cC_1 \oplus_2 \cC_2$ or 
$\cC_1 \d3s \cC_2$ (as the case may be) to $\cC$, 
can be determined in polynomial time. These facts have some significant
consequences, one of which is that any code in $\mfG$ is ML-decodable 
in polynomial time. However, rather than state these results just for 
the class of geometrically perfect codes, we will state and prove them 
more generally for codes that are ``almost graphic'' in the sense
that they can be composed
from graphic codes and finitely many other codes.

Recall that $\G$ denotes the family of graphic codes.

\begin{definition}
A minor-closed family of codes $\mfC$ is defined to be 
\emph{almost-graphic} if there exists a finite 
sub-family $\mfD \subset \mfC$ such that any 2-connected code $\cC \in \mfC$ 
is either in $\G \cup \mfD$, or is of the form $\pi(\cC_1 \oplus_2 \cC_2)$
or $\pi(\cC_1 \d3s \cC_2)$, 
for some permutation $\pi$ of the coordinate set of $\cC$,
and some codes $\cC_1$ and $\cC_2$ such that 
\begin{itemize}
\item[(a)] $\cC_1$ and $\cC_2$ are equivalent to minors of $\cC$, and
\item[(b)] $\cC_1 \in \G \cup \mfD$ and $\cC_2$ is 2-connected.
\end{itemize}

If there exists a constant $l > 0$ such that 
for any length-$n$ code $\cC \in \mfC - (\G \cup \mfD)$, the 
components $\cC_1$, $\cC_2$ and $\pi$ of the above decomposition can 
be determined in time $O(n^l)$, then the family of codes $\mfC$ is said to be 
\emph{polynomially almost-graphic (PAG)}.
\label{almost_graphic_def}
\end{definition}

Note that the 3-sum is conspicuous by its absence from the above definition.
We will give an explanation of this at the end of this section.

Definition~\ref{almost_graphic_def} clearly implies
that any code in an almost-graphic family 
$\mfC$ has a $\bar{3}$-homogeneous code-decomposition tree. The definition
in fact implies that a 2-connected code in $\mfC$ has a decomposition 
tree with the property that each leaf is in $\G \cup \mfD$, and 
for each non-leaf node \sfv in the tree, 
\textsf{lchild.code} is a leaf (and hence, is in $\G \cup \mfD$), 
where \textsf{lchild} is the left child of \textsf{v}. 
Such a code-decomposition tree will be called \emph{$(\G \cup \mfD)$-unary}. 
If $\mfC$ is PAG, a $\bar{3}$-homogeneous, $(\G \cup \mfD)$-unary decomposition
tree can be constructed for any 2-connected code $\cC \in \mfC$ in time 
polynomial in the length of $\cC$.


The family, $\G$, of graphic codes is trivially PAG. From the
discussion prior to the above definition, 
the family, $\mfG$, of geometrically perfect codes
is also PAG. Other examples of PAG families are the 
code families $\mfC_\cF$ (cf.\ Section~\ref{minor_closed_section})
for $\cF = \{\cH_7,\cC(K_5)^\perp\}$ and 
$\cF = \{\cH_7^\perp,\cC(K_5)^\perp\}$, and the 
family of co-graphic codes without a $\cC(K_5)^\perp$ minor 
\cite[Corollary~6.6]{GT}.
PAG codes inherit some of the properties of graphic codes.
For example, it is known \cite{NH81},\cite{jungnickel} 
that the ML decoding problem over a memoryless binary symmetric channel 
can be solved in polynomial time for the family of graphic codes using 
Edmonds' matching algorithm \cite{E2},\cite{EJ73}. A much stronger 
decoding result can in fact be proved for graphic codes, and more generally
for PAG codes. This is based on the following optimization result
proved in \cite{GT}, an argument for which is sketched 
in Appendix~\ref{opt_app}.

\begin{theorem}[\cite{GT}, Theorem~6.5]
Let $\mfC$ be a PAG family of codes. There exists a constant $l > 0$
such that given any length-$n$ code $\cC$ in $\mfC$ and
any $\mathbf{\gm} \in \R^n$, a codeword
$\c_{\min} \in \cC$ achieving $\min_{\c \in \cC} \la\mathbf{\gm},\c\ra$
(or equivalently, $\min_{\x \in P(\cC)} \la\mathbf{\gm},\x\ra$)
can be determined in $O(n^l)$ time.
\label{PAG_opt_thm}
\end{theorem}

It should be noted that an actual implementation of the polynomial-time
algorithm implicit in Theorem~\ref{PAG_opt_thm} (and outlined in 
Appendix~\ref{opt_app}) requires arithmetic over
the real numbers, unless the vector $\gm$ has only rational coordinates.
So, in practice, finite-precision arithmetic used in any computer
implementation of the algorithm could only approximate the linear 
cost function $\la\mathbf{\gm},\c\ra$ for an arbitrary $\gm \in \R^n$.

As mentioned earlier, ML decoding over a discrete memoryless channel can 
be formulated as a linear program \cite{FWK}. Therefore, we have the 
following corollary to the above theorem, again with the caveat 
that a true implementation of a polynomial-time algorithm for ML decoding
would require real-number arithmetic.

\begin{corollary}
The maximum-likelihood decoding problem over a discrete memoryless channel
can be solved in polynomial time for a PAG family of codes. 
In particular, geometrically perfect codes are ML-decodable in polynomial time.
\label{PAG_cor1}
\end{corollary}

Another problem that is known to be NP-hard in general is the problem 
of determining the minimum distance of a code \cite{vardy97}. For 
graphic codes, this is equivalent to the problem of finding the girth of 
a graph, which can be solved in time polynomial in the number of edges
of the graph --- one of the earliest such algorithms published \cite{IR78} 
runs in $O(n^{3/2})$ time in the worst case, where $n$ is the number of edges.
As a consequence of Theorem~\ref{PAG_opt_thm}, we also have that 
the minimum distance problem for a PAG family of codes can be solved
in polynomial time.

\begin{corollary}
The minimum distance of any code in a PAG family can be determined
in polynomial time.
\label{PAG_cor2}
\end{corollary}
\begin{proof}
Let $\cC$ be a code of length $n$ containing at least one nonzero codeword. 
For $i = 1,2,\ldots,n$, define $\mathbf{\gm}^{(i)} = 
(1,\ldots,1,-n,1,\ldots,1)$, with the $-n$ appearing
in the $i$th coordinate of $\mathbf{\gm}^{(i)}$. 
Note that $\min_{\c \in \cC} \la\mathbf{\gm}^{(i)},\c\ra$ is always
achieved by a codeword in $\cC$ with $i$th coordinate equal to 1, if such
a codeword exists. Indeed, the minimum-achieving codeword $\c^{(i)}$
is the codeword of least Hamming weight among codewords in $\cC$ 
that have a 1 in the $i$th coordinate; if there is no codeword in $\cC$ 
with a 1 in the $i$th coordinate, then $\c^{(i)} = \0$. Note that
if $\c^{(i)} \neq \0$, then $\la\mathbf{\gm}^{(i)},\c^{(i)}\ra = 
w(\c^{(i)}) - 1 - n$. Therefore, the minimum distance of $\cC$ is given by 
$$
d = n+1 + \min_{i \in [n]} \min_{\c \in \cC} \la\mathbf{\gm}^{(i)},\c\ra.
$$

For a PAG family of codes $\mfC$, given any code $\cC \in \mfC$, 
each of the minimization problems 
$\min_{\c \in \cC} \la\mathbf{\gm}^{(i)},\c\ra$ can be solved in polynomial
time, and hence the minimum distance of $\cC$ 
can be determined in polynomial time.
\end{proof}

We point out that in all the minimization problems that must be 
solved (recursively going down the code-decomposition tree, as explained
in Appendix~\ref{opt_app}) 
to determine the minimum distance of a length-$n$ code, 
the cost vectors have integer coefficients of magnitude at most
$n^2$. So, the minimum distance of a code from a PAG family can be
determined in polynomial time using finite-precision arithmetic.

The downside of PAG (and more generally, almost-graphic) 
code families is that they are not very good from
a coding-theoretic perspective. Recall from coding theory that a code 
family $\mfC$ is called \emph{asymptotically good} if there exists a
sequence of $[n_i,k_i,d_i]$ codes $\cC_i \in \mfC$, with 
$\lim_i n_i = \infty$, such that $\liminf_i k_i/n_i$ and 
$\liminf_i d_i/n_i$ are both strictly positive.

\begin{theorem}
An almost-graphic family of codes cannot be asymptotically good.
\label{AG_bad_thm}
\end{theorem}

In particular, the family of geometrically perfect codes is not
asymptotically good. The theorem is proved in Appendix~\ref{AG_bad_app}.
In view of Lemma~\ref{pseudocodeword_lemma}, the above result
has the following very interesting corollary.

\begin{corollary}
Let $\mfC$ be a family of binary linear codes with the following
property: for each $\cC \in \mfC$, there exists a parity-check matrix
$H$ for $\cC$, such that the corresponding fundamental polytope $Q(H)$
has no non-integral vertices (pseudocodewords). Then, $\mfC$
is not an asymptotically good code family.
\end{corollary}

Loosely speaking, this means that linear-programming decoding, when applied
to a ``good'' code, must suffer on occasion from decoding failure 
due to the presence of pseudocodewords, even if \emph{all} 
possible parity checks (dual codewords) are used in the constraints 
(\ref{Q_eq1})--(\ref{Q_eq2}) of the LP. Given the close
relationship between linear-programming decoding and iterative decoding
using the min-sum algorithm \cite{VK04}, 
a similar result is likely to hold for iterative decoding as well.

\medskip

We end this section with an explanation of why 3-sums
were left out of Definition~\ref{almost_graphic_def}. The proof of 
Theorem~\ref{PAG_opt_thm} given in Appendix~\ref{opt_app} relies crucially
on the fact that a $\bar{3}$-homogeneous, $(\G \cup \mfD)$-unary 
code-decomposition tree can be constructed in polynomial time for
a code from a PAG family. So, the result is actually 
true for any code family for which such trees can be constructed in 
polynomial time. Now, if 3-sums were allowed in 
Definition~\ref{almost_graphic_def}, it is no longer obvious that 
codes from the resulting code family would still have 
$\bar{3}$-homogeneous, $(\G \cup \mfD)$-unary code-decomposition trees.
There is good reason to think that this could still be true,
especially in light of Corollary~\ref{3sum_cor},
which states that a code has a 3-sum decomposition only if it has 
a $\bar{3}$-sum decomposition. Indeed, that result may lead us to believe
that if a code $\cC$ has a $(\G \cup \mfD)$-unary code-decomposition tree
which contains 3-sums, then replacing the 3-sums in the tree 
with $\bar{3}$-sums in the manner prescribed by Corollary~\ref{3sum_cor} 
should result in a $\bar{3}$-homogeneous, $(\G \cup \mfD)$-unary 
code-decomposition tree. However, to show that this is the case,
it would have to be verified that if $\cC = \cC_1 \oplus_3 \cC_2$, with 
$\cC_1$, $\cC_2$ satisfying (a) and (b) of Definition~\ref{almost_graphic_def},
then (in the notation of Corollary~\ref{3sum_cor}) $\bar{\cC_1}$ and 
$\bar{\cC_2}$ also satisfy (a) and (b). Now, it is not hard to check that 
if $\cC_1$ is graphic, then so is $\bar{\cC_1}$. However, unless
$\cC$ is 3-connected, there is no guarantee that $\bar{\cC_1}$ and 
$\bar{\cC_2}$ are equivalent to minors of $\cC$, even though 
$\cC_1$ and $\cC_2$ are given to be equivalent to minors of $\cC$. 

It is in fact quite likely to be true that if $\cC = \cC_1 \oplus_3 \cC_2$,
with $\cC_1$ and $\cC_2$ equivalent to minors of $\cC$, 
then $\bar{\cC_1}$ and $\bar{\cC_2}$ are also equivalent to minors of $\cC$.
So, it is quite possible that if Definition~\ref{almost_graphic_def}
were to include 3-sums as well, then the resulting code families
would still have $\bar{3}$-homogeneous, $(\G \cup \mfD)$-unary 
code-decomposition trees, and hence Theorem~\ref{PAG_opt_thm} would 
continue to hold. However, a rigorous proof of this would take us far
outside the main theme of our paper, and Definition~\ref{almost_graphic_def}
as it stands is good enough for our purposes.

\section{Concluding Remarks\label{conclusion}}

A natural question to ask upon studying the decomposition theory
presented in this paper is whether one can define $k$-sums for $k \geq 4$
that have the same attractive properties as 2-, 3- and $\bar{3}$-sums. 
Ideally, such a $k$-sum, denoted by $\oplus_k$, would have the 
following property for some fixed integer $l \geq k$: 
\begin{quote}
a $k$-connected code $\cC$ has a $k$-separation $(J,J^c)$
with $\min\{|J|,|J^c|\} \geq l$ iff 
$\cC = \pi(\cC_1 \oplus_k \cC_2)$ for some permutation $\pi$ of
the coordinates of $\cC$, and codes $\cC_1$ and $\cC_2$ equivalent to
minors of $\cC$.
\end{quote}

It is indeed possible to define $k$-sums in such a way that we have
$\cC = \pi(\cC_1 \oplus_k \cC_2)$ iff $\cC$ has a $k$-separation $(J,J^c)$
with $\min\{|J|,|J^c|\} \geq l$. The tricky part is ensuring that
the component codes $\cC_1$ and $\cC_2$ are retained as minors of $\cC$,
and this appears to be difficult in general. 
However, we have some preliminary results that indicate 
that even without the last property, such $k$-sums can be 
used as the building blocks of a decomposition theory that ties in 
beautifully with Forney's theory of cycle-free realizations 
of linear codes \cite{For03}. This theory would make
further deep connections with matroid theory,
particularly with the notions of matroid branchwidth
and treewidth \cite{GGW02}, \cite{hlineny}. 
An exposition of this theory will be given in a future paper.

While the decomposition theory in this paper has been presented 
mainly in the context of binary linear codes, 
it is possible to extend some of it 
to linear codes over arbitrary finite fields as well. 
The definitions of minors and $k$-connectedness can be 
obviously extended to nonbinary codes.
Also, a 2-sum operation can be defined for codes over an 
arbitrary finite field $\F$, and the entire theory 
outlined in Section~\ref{2sum_section} does carry over.
However, there is no known 3-sum operation defined for 
nonbinary codes that has a property analogous to that
stated in Theorem~\ref{3sum_thm}. Again, it seems that
if we are prepared to give up the requirement that the 
components of a $k$-sum be retained as minors of the composite code,
then it is possible to develop a powerful decomposition theory for
nonbinary codes just like that for binary codes.

We end with a pointer to a very interesting direction of current 
research in  matroid theory. This involves the resolution 
of two conjectures whose statements (in the context of codes) 
we give below. Recall from Section~\ref{minor_closed_section} 
that the notation $\mfC_\cF$, for some fixed collection $\cF$ of codes, 
refers to the set of all binary linear codes $\cC$ such that no 
minor of $\cC$ is equivalent to any $\cC' \in \cF$. We extend that notation
to codes over an arbitrary finite field $\F$ as well.
\begin{conjecture}[\cite{GGW_survey}, Conjecture~1.2]
If $\mfC$ is a minor-closed class of codes over a finite field $\F$,
then $\mfC = \mfC_\cF$ for some finite collection of codes $\cF$.
\end{conjecture}
Informally, the above conjecture states that any minor-closed class 
of codes is characterized by a finite list of excluded minors.

\begin{conjecture}[\cite{GGW_survey}, Conjecture~1.3]
Let $\cM$ be a fixed code. Given a length-$n$ code $\cC$, it is decidable
in time polynomial in $n$ whether or not $\cC$ contains $\cM$ as a minor.
\end{conjecture}

The two conjectures together imply that the membership of a code in a
minor-closed class can always be decided in polynomial time. To put it 
another way, if a property of codes is preserved under the action
of taking minors, then it should be decidable in polynomial time
whether or not a given code has that property. It should be pointed
out that both conjectures have been shown to be true in the context
of graphic codes, as part of the celebrated Graph Minor Project 
of Robertson and Seymour \cite{RS13},\cite{RS20}.
The Graph Minor Project has had a profound impact on modern graph theory
\cite{lovasz}, and its extension to $\F$-representable matroids
(equivalently, codes over $\F$) is bound to have a similar influence
on matroid theory and, as a consequence, on coding theory.

\appendix

\section{Proofs of Propositions~\ref{2sum_prop2} 
and \ref{3sum_prop2}\label{dual_props_app}}

\emph{Proof of Proposition~\ref{2sum_prop2}\/}:
Let $\cC$ and $\cC'$ be $[n,k]$ and $[n',k']$ codes, respectively,
for which $\cC \oplus_2 \cC'$ can be defined.
We want to show that 
${(\cC \oplus_2 \cC')}^\perp = \cC^\perp \oplus_2 {\cC'}^\perp$.

First, observe that, by Proposition~\ref{2sum_prop1}(a),
\begin{eqnarray*}
\dim({(\cC \oplus_2 \cC')}^\perp) 
 &=& (n + n' - 2) - \dim(\cC \oplus_2 \cC') 
\ \ = \ \ (n + n' - 2) - (k + k' - 1) \\
 &=& (n-k) + (n'-k') - 1 
\ \ = \ \ \dim(\cC^\perp \oplus_2 {\cC'}^\perp).
\end{eqnarray*}
Therefore, it is enough to show that 
$\cC^\perp \oplus_2 {\cC'}^\perp \subset {(\cC \oplus_2 \cC')}^\perp $.

Given a binary word $\x = x_1 x_2 \ldots x_n$ and a positive 
integer $m \leq n$, let $\x_{:m}$ denote the length-$m$ prefix of $\x$, 
and let $\x_{m:}$ denote the length-$m$ suffix of $\x$.
Also, we denote the concatenation of two binary words $\x$ and $\x'$
by $\x \pl \x'$.

Consider an arbitrary codeword, $\bar{\x}$, 
of $\cC^\perp \oplus_2 {\cC'}^\perp$. 
Such a word is of the form $\x_{[:n-1]} \pl \x'_{[n'-1:]}$ 
for some $\x=(x_1,\ldots,x_n) \in \cC^\perp$ and 
$\x'=(x'_1,\ldots,x'_{n'}) \in {\cC'}^\perp$ such that
$x_n = x'_1$. We will show that
$\x_{[:n-1]} \pl \x'_{[n-1:]}$ as above must also be in 
${(\cC \oplus_2 \cC')}^\perp$. 

Let $\bar{\c}$ be an arbitrary codeword of ${(\cC \oplus_2 \cC')}$.
Such a codeword is of the form $\c_{[:n-1]} \pl \c'_{[n'-1:]}$ 
for some $\c=(c_1,\ldots,c_n) \in \cC$ and 
$\c'=(c'_1,\ldots,c'_{n'}) \in \cC'$ such that
$c_n = c'_1$. The dot product $\bar{\c} \cdot \bar{\x}$ evaluates to
$\sum_{i=1}^{n-1} c_ix_i + \sum_{i=2}^{n'} c'_ix'_i$, all addition operations
being performed modulo 2. But since $\c \in \cC$ and $\x \in \cC^\perp$,
we have $\sum_{i=1}^{n} c_ix_i = 0$, from which we obtain 
$\sum_{i=1}^{n-1} c_ix_i = c_nx_n$. Similarly, $\sum_{i=2}^{n'} c'_ix'_i
= c'_1x'_1$. Hence, $\bar{\c} \cdot \bar{\x} = c_nx_n + c'_1x'_1 = 0$,
since $c_n = c'_1$ and $x_n = x'_1$. As $\bar{\c}$ is
an arbitrary codeword of ${(\cC \oplus_2 \cC')}$, we have shown
that $\bar{\x} \in {(\cC \oplus_2 \cC')}^\perp$.

Therefore, 
$\cC^\perp \oplus_2 {\cC'}^\perp \subset {(\cC \oplus_2 \cC')}^\perp$,
which completes the proof.
\qed
\mbox{ } \\

The proof of Proposition~\ref{3sum_prop2} closely resembles that
of Proposition~\ref{2sum_prop2}, so we continue to use
the notation introduced in the latter proof.\\

\emph{Proof of Proposition~\ref{3sum_prop2}\/}:
Let $\cC$ and $\cC'$ be $[n,k]$ and $[n',k']$ codes, respectively,
for which $\cC \oplus_3 \cC'$ can be defined. Observe that, by 
Proposition~\ref{3sum_prop1} and Lemma~\ref{3barsum_lemma}, we have
\begin{eqnarray*}
\dim({(\cC \oplus_3 \cC')}^\perp) 
 &=& (n + n' - 6) - \dim(\cC \oplus_3 \cC') 
\ \ = \ \ (n + n' - 6) - (k + k' - 4) \\
 &=& (n-k) + (n'-k') - 2 
\ \ = \ \ \dim(\cC^\perp \d3s {\cC'}^\perp).
\end{eqnarray*}
Therefore, to prove Proposition~\ref{3sum_prop2},
it is enough to show that 
$\cC^\perp \d3s {\cC'}^\perp \subset {(\cC \oplus_3 \cC')}^\perp $.


It is easily seen that an arbitrary codeword, $\widehat{\x}$, 
of $\cC^\perp \d3s {\cC'}^\perp$ 
must be of the form $\x_{[:n-3]} \pl \x'_{[n'-3:]}$ for some 
$\x=(x_1,\ldots,x_n) \in \cC^\perp$ and 
$\x'=(x'_1,\ldots,x'_{n'}) \in {\cC'}^\perp$ 
such that $(x_{n-2},x_{n-1},x_n) = (x'_1,x'_2,x'_3)$. 
We will show that any such $\widehat{\x}$ is also
in ${(\cC \oplus_3 \cC')}^\perp$. 

Let $\widehat{\c}$ be an arbitrary codeword of ${(\cC \oplus_3 \cC')}$,
so that it is of the form $\c_{[:n-3]} \pl \c'_{[n'-3:]}$ 
for some $\c=(c_1,\ldots,c_n) \in \cC$ and 
$\c'=(c'_1,\ldots,c'_{n'}) \in \cC'$ such that
$(c_{n-2},c_{n-1},c_n) = (c'_1,c'_2,c'_3)$. 
The dot product $\widehat{\c} \cdot \widehat{\x}$ evaluates to
$\sum_{i=1}^{n-3} c_ix_i + \sum_{i=4}^{n'} c'_ix'_i$, 
all addition operations being performed modulo 2. 
But since $\c \in \cC$ and $\x \in \cC^\perp$,
we have $\sum_{i=1}^{n} c_ix_i = 0$, from which we obtain 
$\sum_{i=1}^{n-3} c_ix_i = \sum_{i=n-2}^n c_ix_i$. 
Similarly, $\sum_{i=4}^{n'} c'_ix'_i
= \sum_{i=1}^3 c'_ix'_i$. 
Hence, $\widehat{\c} \cdot \widehat{\x} = \sum_{i=n-2}^n c_ix_i 
+ \sum_{i=1}^3 c'_ix'_i$, which equals 0 
since $c_{n+i-3} = c'_i$ and $x_{n+i-3} = x'_i$ for $i=1,2,3$. 
As $\widehat{\c}$ is an arbitrary codeword of ${(\cC \oplus_3 \cC')}$, 
we have shown that $\widehat{\x} \in {(\cC \oplus_3 \cC')}^\perp$.

Therefore, 
$\cC^\perp \d3s {\cC'}^\perp \subset {(\cC \oplus_3 \cC')}^\perp $.
which completes the proof.
\qed

\section{Sketch of Proof of Theorem~\ref{PAG_opt_thm}\label{opt_app}}

We will only provide a sketch of the proof of Theorem~\ref{PAG_opt_thm},
as it is a result extant in the literature \cite[Theorem~6.5]{GT}. The
purpose of sketching out the proof is that it outlines the 
polynomial-time algorithm that determines 
\begin{equation}
\min_{\c \in \cC} \la\mathbf{\gm},\c\ra,
\label{min_prob}
\end{equation}
for a length-$n$ code $\cC$ from a PAG family, and a cost vector 
$\mathbf{\gm} \in \R^n$.

The proof of the theorem is by induction using the following lemma
and the fact (explained further below) that the minimization of 
a linear cost function can be done in polynomial time for graphic codes.

\begin{lemma} \emph{(a)}\ 
Let $\cC = \cC_1 \oplus_2 \cC_2$. Then the minimum in 
(\ref{min_prob}) can be obtained by solving two minimization problems
of the form $\min_{\c \in \cC_1} \la\alpha,\c\ra$ and one problem of
the form $\min_{\c \in \cC_2} \la\beta,\c\ra$. \\
\emph{(b)}\ Let $\cC = \cC_1 \d3s \cC_2$. 
Then the minimum in (\ref{min_prob}) can be obtained by solving four 
minimization problems of the form $\min_{\c \in \cC_1} \la\alpha,\c\ra$ 
and one problem of the form $\min_{\c \in \cC_2} \la\beta,\c\ra$.
\label{min_lemma}
\end{lemma}
\begin{proof}
Let $n$, $n_1$ and $n_2$ denote the lengths of $\cC$, $\cC_1$ and $\cC_2$,
respectively. Given $\gm \in \R^n$, let $M = 1 + \sum_{i=1}^n |\gm_i|$.

(a)\ Let $\cC = \cC_1 \oplus_2 \cC_2$.
Define $\alpha^{(0)} = (\alpha^{(0)}_1, \ldots, \alpha^{(0)}_{n_1})$
and $\alpha^{(1)} = (\alpha^{(1)}_1, \ldots, \alpha^{(1)}_{n_1})$ as follows: 
\begin{eqnarray*}
\alpha^{(0)}_i &=& \left\{
\begin{array}{rl}
\gm_i & \ \ \ i = 1,\ldots,n_1-1 \\
M & \ \ \ i = n_1
\end{array} 
\right. \\
\alpha^{(1)}_i &=& \left\{
\begin{array}{cl}
\gm_i & \ \ \ i = 1,\ldots,n_1-1 \\
-M & \ \ \ i = n_1
\end{array} 
\right. 
\end{eqnarray*}
For $j = 0,1$, determine $\mu_j = \min_{\c \in \cC_1} \la\alpha^{(j)},\c\ra$,
and a minimum-achieving codeword $\c^{(j)} = (c^{(j)}_1,\ldots,c^{(j)}_{n_1})
\in \cC_1$. By choice of $M$, we have that $c^{(0)}_{n_1} = 0$, while 
$c^{(1)}_{n_1} = 1$. 

Now, define $\beta = (\beta_1,\ldots,\beta_{n_2})$ as follows:
\begin{eqnarray*}
\beta_i &=& \left\{
\begin{array}{cl}
\mu_1 - \mu_0 + M  & \ \ \ i = 1 \\
\gm_{n_1+i-2} & \ \ \ i = 2,\ldots,n_2.
\end{array} 
\right.
\end{eqnarray*}
Solve for $\widehat{\mu} = \min_{\c \in \cC_2} \la\beta,\c\ra$, and find
a codeword $\hbc = (\hc_1,\ldots,\hc_{n_2}) \in \cC_2$ achieving this minimum. 

We claim that 
$\min_{\c \in \cC} \la\mathbf{\gm},\c\ra = \mu_0 + \widehat{\mu}$,
and that a minimum-achieving codeword in $\cC$ is 
$$
\c_{\min} = \left\{
\begin{array}{cl}
(c^{(0)}_1,\ldots,c^{(0)}_{n_1-1},\hc_2,\ldots,\hc_{n_2}) 
&\ \ \text{ if } \hc_1 = 0 \\
(c^{(1)}_1,\ldots,c^{(1)}_{n_1-1},\hc_2,\ldots,\hc_{n_2}) 
&\ \ \text{ if } \hc_1 = 1.
\end{array}
\right.
$$
We will first show that for each $\c \in \cC$, 
we have $\la\mathbf{\gm},\c\ra \geq \mu_0 + \widehat{\mu}$.
Pick an arbitrary $\c \in \cC$. There exists a unique pair 
of codewords $\x = (x_1,\ldots,x_{n_1}) \in \cC_1$, 
$\y = (y_1,\ldots,y_{n_2}) \in \cC_2$ such that $x_{n_1} = y_1$,
and $(x_1,\ldots,x_{n_1-1},y_2,\ldots,y_{n_2}) = \c$. Suppose
that $x_{n_1} = y_1 = 0$. We then have 
$$
\la\gm,\c\ra = \la\alpha^{(0)},\x\ra + \la\beta,\y\ra 
\geq \mu_0 + \widehat{\mu}.
$$
Next, suppose that $x_{n_1} = y_1 = 1$. In this case, we have
\begin{eqnarray*}
\la\gm,\c\ra &=& \la\alpha^{(1)},\x\ra - \alpha^{(1)}_{n_1} + 
\la\beta,\y\ra - \beta_1 \\
& = &  \la\alpha^{(1)},\x\ra - (-M) +
\la\beta,\y\ra - (\mu_1-\mu_0 + M)  \\
& \geq & \mu_1 - (-M) + \widehat{\mu} - (\mu_1-\mu_0 + M) 
\ \ =\ \ \mu_0 + \widehat{\mu}.
\end{eqnarray*}
Thus, $\la\mathbf{\gm},\c\ra \geq \mu_0 + \widehat{\mu}$, as desired.

It is now enough to show that 
$\la\mathbf{\gm},\c_{\min} \ra = \mu_0 + \widehat{\mu}$.
By definition of $\c_{\min}$, 
$$
\la\mathbf{\gm},\c_{\min} \ra = \left\{
\begin{array}{cl}
\la\alpha^{(0)},\c^{(0)}\ra + \la\beta,\hbc\ra
&\ \ \text{ if } \hc_1 = 0 \\
\la\alpha^{(1)},\c^{(1)}\ra - (-M) + \la\beta,\hbc\ra - (\mu_1-\mu_0+M)
&\ \ \text{ if } \hc_1 = 1.
\end{array}
\right.
$$
In either case, $\la\mathbf{\gm},\c_{\min} \ra = \mu_0 + \widehat{\mu}$.

(b)\ Let $\cC = \cC_1 \d3s \cC_2$. Note that 
from (A1$'$)--(A3$'$) in Definition~\ref{3barsum_def}, it follows 
that the restriction of $\cC_1$ (resp.\ $\cC_2$) onto its last 
(resp.\ first) three coordinates is $\{000,011,101,110\}$.

Define 
$\alpha^{(j)} = (\alpha^{(j)}_1,\ldots,\alpha^{(j)}_{n_1})$,
$j = 0,1,2,3$, as follows: 
$\alpha^{(j)}_i = \gm_i$ for $i = 1,\ldots,n_1-3$, and 
$$
\begin{array}{rcrcrcc}
\alpha^{(0)}_{n_1-2} & = & \alpha^{(0)}_{n_1-1} & = & \alpha^{(0)}_{n_1} 
& = & M; \\
\alpha^{(1)}_{n_1-2} & = & - \alpha^{(1)}_{n_1-1} & = & - \alpha^{(1)}_{n_1} 
& = & M; \\
- \alpha^{(2)}_{n_1-2} & = & \alpha^{(2)}_{n_1-1} & = & - \alpha^{(2)}_{n_1} 
& = & M; \\
- \alpha^{(3)}_{n_1-2} & = & - \alpha^{(3)}_{n_1-1} & = & \alpha^{(3)}_{n_1} 
& = & M.
\end{array} 
$$
For $j = 0,1,2,3$, determine 
$\mu_j = \min_{\c \in \cC_1} \la\alpha^{(j)},\c\ra$,
and a minimum-achieving codeword 
$\c^{(j)} = (c^{(j)}_1,\ldots,c^{(j)}_{n_1}) \in \cC_1$. 
By choice of the $\alpha^{(j)}$'s, we have that 
$$
(c^{(j)}_{n_1-2},c^{(j)}_{n_1-1},c^{(j)}_{n_1}) = \left\{
\begin{array}{cl}
000 & \ \ \text{ if } j = 0 \\
011 & \ \ \text{ if } j = 1 \\
101 & \ \ \text{ if } j = 2 \\
110 & \ \ \text{ if } j = 3.
\end{array}
\right.
$$

Now, take $\beta = (\beta_1,\ldots,\beta_{n_2})$ to be
\begin{eqnarray*}
\beta_i &=& \left\{
\begin{array}{cl}
- (\mu_0 + \mu_1 - \mu_2 - \mu_3)/2 + M & \ \ \ i = 1 \\
- (\mu_0 - \mu_1 + \mu_2 - \mu_3)/2 + M & \ \ \ i = 2 \\
- (\mu_0 - \mu_1 - \mu_2 + \mu_3)/2 + M & \ \ \ i = 3 \\
\gm_{n_1+i-6} & \ \ \ i = 4,\ldots,n_2.
\end{array} 
\right.
\end{eqnarray*}
Solve for $\widehat{\mu} = \min_{\c \in \cC_2} \la\beta,\c\ra$, and find
a codeword $\hbc = (\hc_1,\ldots,\hc_{n_2}) \in \cC_2$ achieving this minimum. 
It may be verified that 
$\min_{\c \in \cC} \la\mathbf{\gm},\c\ra = \mu_0 + \widehat{\mu}$,
and that a minimum-achieving codeword in $\cC$ is 
$$
\c_{\min} = \left\{
\begin{array}{cl}
(c^{(0)}_1,\ldots,c^{(0)}_{n_1-3},\hc_4,\ldots,\hc_{n_2}) 
&\ \ \text{ if } (\hc_1,\hc_2,\hc_3) = 000 \\
(c^{(1)}_1,\ldots,c^{(1)}_{n_1-3},\hc_4,\ldots,\hc_{n_2}) 
&\ \ \text{ if } (\hc_1,\hc_2,\hc_3) = 011 \\
(c^{(2)}_1,\ldots,c^{(2)}_{n_1-3},\hc_4,\ldots,\hc_{n_2}) 
&\ \ \text{ if } (\hc_1,\hc_2,\hc_3) = 101 \\
(c^{(3)}_1,\ldots,c^{(3)}_{n_1-3},\hc_4,\ldots,\hc_{n_2}) 
&\ \ \text{ if } (\hc_1,\hc_2,\hc_3) = 110.
\end{array}
\right.
$$
The details of the verification are along the lines of that in 
part (a), and are left to the reader.
\end{proof}

Suppose that we have a PAG code family $\mfC$, and must solve (\ref{min_prob})
for a given length-$n$ code $\cC \in \mfC$ and cost vector
$\gm = (\gm_1,\ldots,\gm_n) \in \R^n$. Note that if $\cC$ can be expressed
as a direct sum, $\cC_1 \oplus \cC_2$, of codes $\cC_1$ and $\cC_2$ of 
lengths $n_1$ and $n_2$, respectively, then
$$
\min_{\c \in \cC} \la\gm,\c\ra = 
\min_{\c^{(1)} \in \cC_1} \la\gm^{(1)},\c^{(1)}\ra + 
\min_{\c^{(2)} \in \cC} \la\gm^{(2)},\c^{(2)}\ra,
$$
where $\gm^{(1)} = (\gm_1,\ldots,\gm_{n_1})$ and 
$\gm^{(2)} = (\gm_{n_1+1},\ldots,\gm_{n})$.
Therefore, to show that (\ref{min_prob}) can be solved in time polynomial 
in $n$, it is enough to show that there is a polynomial-time algorithm
in the case when $\cC$ is 2-connected.

It follows from Definition~\ref{almost_graphic_def} that there exists 
a finite sub-family $\mfD \subset \mfC$ such that 
each 2-connected code $\cC \in \mfC$ has a $\bar{3}$-homogeneous, 
$(\G \cup \mfD)$-unary code-decomposition tree, which can be 
constructed in polynomial time. So, an algorithm for solving 
(\ref{min_prob}) for a 2-connected code $\cC \in \mfC$ 
would use Lemma~\ref{min_lemma} to recursively go down the 
code-decomposition tree starting from
the root node, solving at most four minimization problems at each
leaf of the tree. Recall that each leaf of a $(\G \cup {\mfD})$-unary
code-decomposition tree is a code in $\G \cup \mfD$. 
Since such a tree for a length-$n$ code can have at most $n$ leaves, 
the algorithm for solving (\ref{min_prob}) would run in time polynomial 
in $n$, provided that there is a polynomial-time algorithm for 
solving (\ref{min_prob}) for codes $\cC \in \G$, \emph{i.e.}, graphic codes.

Indeed, there exists a polynomial-time algorithm for solving (\ref{min_prob}) 
for graphic codes. Note that the minimization problem $\min\la\gm,\c\ra
= \min\sum_{i=1}^n \gm_ic_i$ over a graphic code $\cC(\cG)$ 
is equivalent to the problem of finding the minimum-weight Eulerian 
subgraph in the graph $\cG$ whose edges $e_i$, 
$i = 1,2,\ldots,n$, are given the weights $\gm_i$. 
An Eulerian subgraph of a graph is a subgraph in which 
each vertex has even degree. 

The Eulerian subgraph problem can be solved as 
follows\footnote{This approach to solving the Eulerian subgraph problem
was conveyed to the author by Adrian Vetta.}. Given a subset $T$ of the
vertices of $\cG$, a \emph{T-join} is a set of edges $J$ of $\cG$
such that a vertex $v$ in $\cG$ has odd degree with respect to (wrt) $J$ 
if and only if $v \in T$. Let $N$ be the set of edges $\{e_i: \gm_i < 0\}$,
and let $T$ be the subset of vertices with odd degree wrt $N$. Define
the graph $\cG'$ to be the graph $\cG$ but with edge-weights 
$\gm'_i = |\gm_i|$, $i = 1,2,\ldots,n$. Find a minimum-weight $T$-join, $J$,
in $\cG'$. The symmetric difference $J \Delta N$ is the required 
minimum-weight Eulerian subgraph of $\cG$. A minimum-weight
$T$-join can be found in polynomial time \cite[Chapter 29]{schrijver},
and so, a minimum-weight Eulerian subgraph of $\cG$ can be determined
in polynomial time.

\section{Proof of Theorem~\ref{AG_bad_thm}\label{AG_bad_app}}

We will first prove that the family, $\G$, of graphic codes is not 
asymptotically good. This is probably a ``folk'' theorem, but we could not
find an explicit proof in the literature. Our proof relies on the 
fact \cite{alon} that for a graph $\cG = (V,E)$ with girth $g$ and 
average degree $\bd = 2|E|/|V| \geq 2$, 
the number of vertices satisfies the so-called \emph{Moore bound}:
\begin{equation*}
|V| \geq \left\{
\begin{array}{cl}
1 + \bd\sum_{i=0}^{\lfloor g/2 \rfloor - 1} (\bd-1)^i & 
\text{if $g$ is odd} \\
2\sum_{i=0}^{\lfloor g/2 \rfloor - 1} (\bd-1)^i & 
\text{if $g$ is even}.
\end{array}
\right.
\end{equation*}
For our purposes, the weaker bound 
$|V| \geq (\bd-1)^{\lfloor g/2 \rfloor - 1}$ is enough, as we then have,
for $\bd > 2$,
\begin{equation}
g \leq 4 + \frac{\log |V|}{\log(\bd-1)} \leq 4 + \frac{\log |E|}{\log(\bd-1)},
\label{g_bnd}
\end{equation}
where, for the sake of concreteness, $\log$ denotes the natural logarithm.

Since codewords in $\cC(\cG)$ correspond to cycles in $\cG$, 
we see that the minimum distance of $\cC(\cG)$ equals the girth $g$
of the graph $\cG$. Furthermore, if $\cG$ is connected, then (as mentioned
in Section~\ref{matroid_section}) the rank of its vertex-edge incidence matrix
is $|V|-1$, and hence, $\dim(\cC(\cG)) = |E| - (|V|-1)$. Thus, the
rate of $\cC(\cG)$ is $1-(|V|-1)/|E| \approx 1-2/\bd$. Consequently,
if a family of graphic codes has dimension growing linearly with 
codelength $n$, then by (\ref{g_bnd}) their minimum distance grows 
as $O(\log n)$, which implies that graphic codes are not asymptotically 
good. This argument is formalized in the proof given below.

\begin{lemma}
The family of graphic codes is not asymptotically good.
\label{graphic_bad_lemma}
\end{lemma}
\begin{proof}
We will in fact prove a stronger statement: for $r \in (0,1)$, let
$\G_r = \{\cC \in \G:\ \text{$\cC$ has rate $> r$}\}$; then,
for any code $\cC \in \G_r$ with length $n \geq 2$, we have 
\begin{equation}
d(\cC) \leq \frac{4 \log n}{\log(1+r/2)}.
\label{dmin_bnd}
\end{equation}

Consider first an $[n,k,d]$ code $\cC \in \G_r$ with $n > 2/r$.
Without loss of generality (WLOG), we can assume that 
$\cC = \cC(\cG)$ for some connected graph $\cG = (V,E)$
\cite[Proposition~1.2.8]{oxley}. Therefore,
$k/n = 1 - (|V|-1)/|E| > r$, or equivalently, 
$(|V|-1)/|E| < 1-r$. Furthermore, since $n > 2/r$, we have that 
$|V|/|E| < 1-r/2$. Therefore, the average degree, $\bd$, of $\cG$
is larger than $2/(1-r/2)$, which is in turn larger than $2+r/2$. 
Hence, by (\ref{g_bnd}), 
$$
d \leq 4 + \frac{\log n}{\log(1+r/2)} \leq \frac{4 \log n}{\log(1+r/2)}.
$$
The last inequality above holds when $\frac{\log n}{\log(1+r/2)}
\geq 4/3$, which is true for $n > 2/r$, since
$\frac{\log (2/r)}{\log(1+r/2)} \geq 4/3$ for $r \in (0,1)$, 
as a simple calculation will confirm. 

Therefore, without the assumption $n > 2/r$, we have
$$
d(\cC) \leq \max\left\{2/r,\ \frac{4 \log n}{\log(1+r/2)}\right\}.
$$
However, it is straightforward to verify that 
$(2/r) \log(1+r/2) \leq 1$ for any $r > 0$, 
from which we obtain that for $n \geq 2$, 
$\frac{4 \log n}{\log(1+r/2)} > 2/r$, and (\ref{dmin_bnd}) follows.
\end{proof}

Given an almost-graphic code family $\mfC$, let 
$\mfD$ be the finite sub-family of codes with the 
properties guaranteed by Definition~\ref{almost_graphic_def}. 
WLOG, we may assume that $\mfD$ is minor-closed.

For $r \in (0,1)$, define $N_r$ to be the least positive
integer such that for all $n > N_r$, 
$$
0 < \frac{1}{\log(1+(r-2/n)/2)} < \frac{2}{\log(1+r/2)}.
$$ 
Note that since
$\lim_{n\rightarrow \infty} 1/\log(1+(r-2/n)/2) = 1/\log(1+r/2)$, 
such an $N_r$ does exist. Now, define
$$
d_{\max}(r,\mfD) = \max\{d(\cC):\ \cC \in \mfC \text{ and has length at most $N_r$,
or $\cC \in \mfD$}\}.
$$

We now have the definitions needed to state the next result, which shows that
codes in $\mfC$ cannot have both dimension and minimum distance growing
linearly with codelength. It is clear
that Theorem~\ref{AG_bad_thm} follows directly from this result.

\begin{lemma}
Let $\mfC$ be an almost-graphic family of codes. 
For any $r \in (0,1)$, if $\cC \in \mfC$ is an $[n,k,d]$ code 
with $k/n > r$, then 
\begin{equation}
d \leq \max\left\{d_{\max}(r,\mfD),\ \frac{8 \log n}{\log(1+r/2)} \right\}.
\label{AG_dmin_bnd}
\end{equation}
\label{AG_bad_lemma}
\end{lemma}


\begin{proof}
From the definition of $d_{\max}(r,\mfD)$ and (\ref{dmin_bnd}),
it is obvious that the statement of the lemma 
holds for all codes in $\G \cup \mfD$.
The proof that the statement holds for all codes in $\mfC$
is by induction on codelength for a fixed $r \in (0,1)$. 

So, fix an $r \in (0,1)$. If $n_0$ is the smallest length of a 
non-trivial code in $\mfC$, then a length-$n_0$ code in $\mfC$ cannot 
be decomposed into smaller codes, and so must be in $\G \cup \mfD$.
Therefore, the statement of the lemma
holds for the base case of length-$n_0$ codes.

Now, suppose that for some $n > n_0$,
(\ref{AG_dmin_bnd}) holds for all codes $\cC' \in \mfC$ of length 
$n' \leq n-1$ and rate larger than $r$.
Let $\cC \in \mfC$ be an $[n,k,d]$ code with $k/n > r$. 
If $\cC = \cC_1 \oplus \cC_2$ for some codes $\cC_1$ and $\cC_2$ in $\mfC$, 
then at least one of $\cC_1$ and $\cC_2$ has rate larger than $r$,
and so (\ref{AG_dmin_bnd}) holds for $\cC$ by the induction hypothesis.

We may thus assume that $\cC$ is 2-connected.
If $\cC \in \G \cup \mfD$, there is nothing to be proved. So, either
$\cC = \pi(\cC_1 \oplus_2 \cC_2)$ or $\pi(\cC_1 \d3s \cC_2)$ for
$\cC_1,\cC_2,\pi$ as in Definition~\ref{almost_graphic_def}. WLOG, 
we may take $\pi$ to be the identity permutation,
so that $\cC$ is either $\cC_1 \oplus_2 \cC_2$ or $\cC_1 \d3s \cC_2$,
for some $[n_1,k_1]$ code $\cC_1 \in \G \cup \mfD$ and some $[n_2,k_2]$ 
code $\cC_2 \in \mfC$.

We consider the case $\cC = \cC_1 \oplus_2 \cC_2$ first. 
By Proposition~\ref{2sum_prop1}, $k = k_1 + k_2 - 1$, and 
$d \leq \min\{d(\cC_1'),d(\cC_2')\}$, where 
$\cC_1' = \cC_1 \shorten \{n_1\}$ and $\cC_2' = \cC_2 \shorten \{1\}$.
Note that for $i=1,2$, $\cC_i'$ is an $[n_i',k_i']$ code, where
$n_i' = n_i - 1$ and $k_i' = k_i - 1$. Thus, $k = k_1'+k_2'+1$
and $n = n_1'+n_2'$. 

Now, if $k_i'/n_i' > r$ for some $i \in \{1,2\}$, then the
statement of the lemma holds for $\cC$ by the induction hypothesis.
So, we are left with the situation when 
$k_i'/n_i' \leq r$ for $i=1,2$. But in this case, since 
$k/n = (k_1'+k_2'+1)/(n_1'+n_2') > r$, we must have $k_1'/n_1' > r - 1/n_1'$;
otherwise, we would have $k_2' > (n_1'+n_2')r - 1 - k_1' \geq 
(n_1'+n_2')r - 1 - (rn_1'-1) = rn_2'$, which would mean that $k_2'/n_2' > r$.
Note that since $\cC_1 \in \G \cup \mfD$, and $\cC_1'$ is a minor
of $\cC_1$, we have that $\cC_1' \in \G \cup \mfD$. 
If $\cC_1' \in \mfD$ or $n_1' \leq N_r$, 
then $d(\cC_1') \leq d_{\max}(r,\mfD)$;
otherwise, $\cC_1'$ is graphic with $n_1' > N_r$, 
and so, by (\ref{dmin_bnd}) and the definition of $N_r$,
$$
d(\cC_1') \leq \frac{4 \log n_1'}{\log(1+(r-1/n_1')/2)} 
\leq \frac{4 \log n_1'}{\log(1+(r-2/n_1')/2)} \leq \frac{8 \log n_1'}{1+r/2}.
$$
In any case, $d(\cC_1') 
\leq \max\left\{d_{\max}(r,\mfD),\ \frac{8 \log n_1'}{\log(1+r/2)} \right\}
\leq \max\left\{d_{\max}(r,\mfD),\ \frac{8 \log n}{\log(1+r/2)} \right\}$. 
Since $d \leq d(\cC_1')$, we have that (\ref{AG_dmin_bnd}) holds for $\cC$.

Finally, we deal with the case when $\cC = \cC_1 \d3s \cC_2$. The approach
is essentially the same as that in the 2-sum case. This time, defining
$\cC_1' = \cC_1 \shorten \{n_1-2,n_1-1,n_1\}$ and
$\cC_2' = \cC_2 \shorten \{1,2,3\}$, we have 
$d \leq \min\{d(\cC_1'),d(\cC_2')\}$. For $i=1,2$, we now find that
$\cC_i'$ is an $[n_i',k_i']$ code, where
$n_i' = n_i - 3$ and $k_i' = k_i - 2$. Thus, $n = n_1'+n_2'$, and 
via Lemma~\ref{3barsum_lemma}, $k = k_1'+k_2'+2$. If either 
$k_1'/n_1'$ or $k_2'/n_2'$ is larger than $r$, then (\ref{AG_dmin_bnd})
holds for $\cC$ by the induction hypothesis. So suppose that
$k_i'/n_i' \leq r$ for $i = 1,2$. Since $k/n = 
(k_1'+k_2'+1)/(n_1'+n_2') > r$, we must have $k_1'/n_1' > r - 2/n_1'$;
otherwise, we would obtain $k_2'/n_2' > r$. 
If $\cC_1' \in \mfD$ or $n_1' \leq N_r$, 
then $d(\cC_1') \leq d_{\max}(r,\mfD)$;
otherwise, $\cC_1'$ is graphic with $n_1' > N_r$, 
and so, by (\ref{dmin_bnd}) and the definition of $N_r$,
$$
d(\cC_1') \leq \frac{4 \log n_1'}{\log(1+(r-2/n_1')/2)} 
\leq \frac{8 \log n_1'}{1+r/2}.
$$
In any case, $d(\cC_1') \leq 
\max\left\{d_{\max}(r,\mfD),\ \frac{8 \log n}{\log(1+r/2)} \right\}$, and
since $d \leq d(\cC_1')$, we see that (\ref{AG_dmin_bnd}) holds for $\cC$.
The proof of the lemma is now complete.
\end{proof}

\section*{Acknowledgments}
The author would like to thank Jim Geelen for patiently answering
the author's questions pertaining to the structure theory of matroids, 
Adrian Vetta for the polynomial-time algorithm 
for solving the minimum-weight Eulerian subgraph problem,
and Alexander Vardy for discussions leading up to this work. 
Finally, the author is grateful to the anonymous reviewers for
suggesting several improvements to the presentation.

\end{document}